\documentclass[sd, pop, preprint, superscriptaddress]{revtex4-1}
\usepackage{color}
\usepackage{amsmath}
\usepackage{amsfonts}
\usepackage{amssymb}
\usepackage{graphicx}
\usepackage[normalem]{ulem}

\begin{document}


\title{Increased efficiency of ion acceleration by using femtosecond laser pulses at higher harmonic frequency} 



\author{J. Psikal}
\email[]{jan.psikal@fjfi.cvut.cz}
\affiliation{FNSPE, Czech Technical University in Prague, Czech Republic}
\author{O. Klimo}
\affiliation{FNSPE, Czech Technical University in Prague, Czech Republic}
\affiliation{ELI-Beamlines project, Institute of Physics, ASCR, Czech Republic}
\author{S. Weber}
\affiliation{ELI-Beamlines project, Institute of Physics, ASCR, Czech Republic}
\author{D. Margarone}
\affiliation{ELI-Beamlines project, Institute of Physics, ASCR, Czech Republic}


\date{\today}

\begin{abstract}
The influence of laser frequency on laser-driven ion acceleration is investigated by means of two-dimensional particle-in-cell simulations. 
When ultrashort intense laser pulse at higher harmonic frequency irradiates a thin solid foil, the target may become relativistically transparent for significantly lower laser pulse intensity compared to irradiation at fundamental laser frequency. 
The relativistically induced transparency results in an enhanced heating of hot electrons as well as increased maximum energies of accelerated ions and their numbers. 
Our simulation results have shown the increase of maximum proton energy and of the number of high-energy protons by a factor of 2 after the interaction of an ultrashort laser pulse of maximum intensity $7 \times 10^{21}~\rm{W/cm^2}$ with a fully ionized plastic foil of realistic density and of optimal thickness between $100~\rm{nm}$ and $200~\rm{nm}$ when switching from the fundamental frequency to the third harmonics. 
\end{abstract}


\maketitle 

\section{Introduction}
The generation of highly energetic ion beams from laser-plasma interaction has attracted great interest in the last decade \cite{Macchi2013} due to the broad range of applications including cancer therapy \cite{Cirrone2013}, short-lived isotope production for medical applications \cite{Ledingham2010}, isochoric heating of solid-density matter \cite{Snavely2007}, proton radiography \cite{Borghesi2004}, and fast ignition in inertial confinement fusion \cite{Roth2001}. 
Up to now the highest energy ions ($160~\rm{MeV}$ for protons \cite{Hegelich2013a} and $1~\rm{GeV}$ for carbon ions \cite{Jung2013a}) have been achieved with high energy (hundreds of Joules) and relatively long (0.5-1 ps) laser pulses on the Trident laser in the so-called Break Out Afterburner (BOA) regime \cite{Yin2006, Albright2007, Yin2007, Yin2011}. 
Such source of high energy protons and ions is still not useful for societal applications because of the large size of currently used laser installations and the limited repetition rate. 
Nevertheless, smaller laser facilities, delivering a few tens of J and short laser pulses, which can in principle operate at 10 Hz repetition rate \cite{Rus2013} are more promising. 

The BOA regime, where the energy of accelerated protons is currently the highest in comparison with all other experimentally investigated regimes, relies on the fact that the laser pulse can burn through the target. 
In the experiments with longer and relatively high energy pulses, this is achieved naturally due to rapid expansion of the target heated by the first part of the laser pulse \cite{Hegelich2013b, Jung2013b}. 
With shorter laser pulses, this can be however hardly achieved. 
In principle, one may use such an intense pulse that the target becomes relativistically transparent. 
However, such regime is not accessible at the moment. 
According to the estimate provided by D. Jung \emph{et al.} \cite{Jung2013b} based on the previous model \cite{Yan2010}, the onset of the relativistic transparency takes place at the time $t_1=(12/\pi^2)^{1/4}\sqrt{(n_e/n_c) \tau d / (a_0 c_s) }$, where $n_e/n_c$ is the ratio of initial electron density to the critical electron density, $d$ is the target thickness, $a_0$ normalized laser amplitude, $\tau$ laser pulse duration at FWHM, and ion sound speed $c_s \approx \sqrt{Z m_e c^2 a_0 / m_i}$.
The lowest density compact solid matter, which is routinely available (plastic targets) has a free electron density of about 200 $n_c$ (for the wavelength of Ti:Sapphire laser equal to $800~\rm{nm}$), when fully ionized. 
Thus, we can estimate that, for example, for laser pulse duration (FWHM) of $20~\rm{fs}$ and $200~\rm{nm}$ thick plastic foil, laser intensity about $4 \times 10^{22}~\rm{W/cm^2}$ is required to obtain relativistic transparency during laser-target interaction.
Moreover, an ultraintense pulse works partially like a piston pushing a cloud of electrons ahead \cite{Sentoku2003}, which further increases the required intensity for very short pulses \cite{Qiao2012}. 

Nevertheless, there might be another approach how to make the target relativistically transparent than increasing laser pulse energy and the pulse length - converting the laser pulse to higher frequency. 
For example, since the critical electron density $n_c \sim \omega^2$, the same free electron density $n_e$ expressed in terms of critical density is reduced 9 times (e.g., it is reduced to $\sim 20~n_c$ for plastic foil) for the third harmonics compared with the fundamental frequency. 
Obviously, the parameter $a_0$ is also reduced but only 3 times for the third harmonics $(a_0 \sim \sqrt{I} / \omega)$ \cite{Gibbon2005}. 
Since the relativistically induced transparency takes place when $n_c \lesssim n_e \lesssim (1 + a_0^2/2)^{1/2} n_c$ \cite{Lefebvre1995, Sakagami1996}, the intensity required for relativistic transparency can be significantly reduced. 
Namely, $\left(\frac{n_e}{10^{23}~\rm{cm^{-3}}} \times \frac{\lambda}{\rm{\mu m}} \right)^2 \lesssim \frac{I}{2.2 \cdot 10^{22}~\rm{W/cm^{-2}}}$ is required for the target to be relativistically transparent assuming that $n_e > n_c$. 
Again, required laser intensity can be partially reduced by assuming target expansion \cite{Yan2010, Jung2013b}, which is important especially for longer laser pulses and for the similarity parameter \cite{Gordienko2005} $S=n_e/(a_0 n_c)>>1$.

Obvious argument against this approach can be that conversion to the third harmonics costs a lot of energy. 
On the other hand, it is known that this conversion greatly improves the contrast of the laser pulse and not only in the nanosecond domain but also in the picosecond domain \cite{Hillier2013}. 
Thus, the improvement of the laser pulse intensity contrast usually performed through the double plasma mirror technique \cite{Thaury2007}, which implies a similar energy loss, can be avoided in such case. 
Another clear argument can be that the electrons are less heated with lower $a_0$ and thus the proton energy should be smaller. 
However, we will demonstrate in this paper that such point of view does not apply here, because the regime of laser target interaction and proton acceleration is substantially different. 

\section{Comparison of ion acceleration by using fundamental and the third harmonic frequency}
In order to show in more detail the mechanism which was discussed above, we have employed our 2D3V particle-in-cell (PIC) code \cite{Psikal2006} (with two spatial and three velocity components). 
In the simulations, we assumed the interaction of laser pulses with homogeneous fully ionized polyethylene $\rm{CH_2}$ foils of the density $0.9~\rm{g/cm^3}$, where the free electron density is $n_e=3.5 \cdot 10^{23}~\rm{cm^{-3}}$, at the wavelengths $\lambda_1=800~\rm{nm}$ corresponding to the fundamental frequency ($1 \omega$) and $\lambda_3=264~\rm{nm}$ corresponding to the third harmonic frequency ($3 \omega$). 
The electron density $n_e$ is equal to $200~n_c$ and $21.8~n_c$ for $1 \omega$ and $3 \omega$ cases, respectively. 
Ultrahigh laser pulse contrast is assumed, thus, all targets are initialized with step-like density profile. 
In order to prevent numerical heating \cite{Ueda1994}, the initial plasma temperature is set to $3~\rm{keV}$ and the cell size is $8~\rm{nm}$. 
The laser pulse has $\sin^2$ temporal shape of full duration about $40~\rm{fs}$ ($15 \tau$ for $1 \omega$ case and $45 \tau$ for $3 \omega$, where $\tau$ is the laser wave period). 
The peak intensity of linearly (p-)polarized pulse is set to $7.2 \cdot 10^{21}~\rm{W/cm^2}$ (dimensionless amplitudes $a_0=58$ and $a_0=19.1$ for $1 \omega$ and $3 \omega$ cases, respectively, where $I \lambda^2=a_0^2 \times 1.37 \cdot 10^{18}~\rm{[W \cdot \mu m^2 / cm^2]}$). 
The focal spot diameter is set to $3~\rm{\mu m}$ at FWHM (gaussian shape). 
The foils are irradiated at normal incidence since the acceleration of ions has been shown to be more efficient at normal than at oblique incidence for the laser pulse with the same parameters \cite{Limpouch2013}. 

\begin{figure}
\begin{center}
\includegraphics[width=0.45\textwidth]{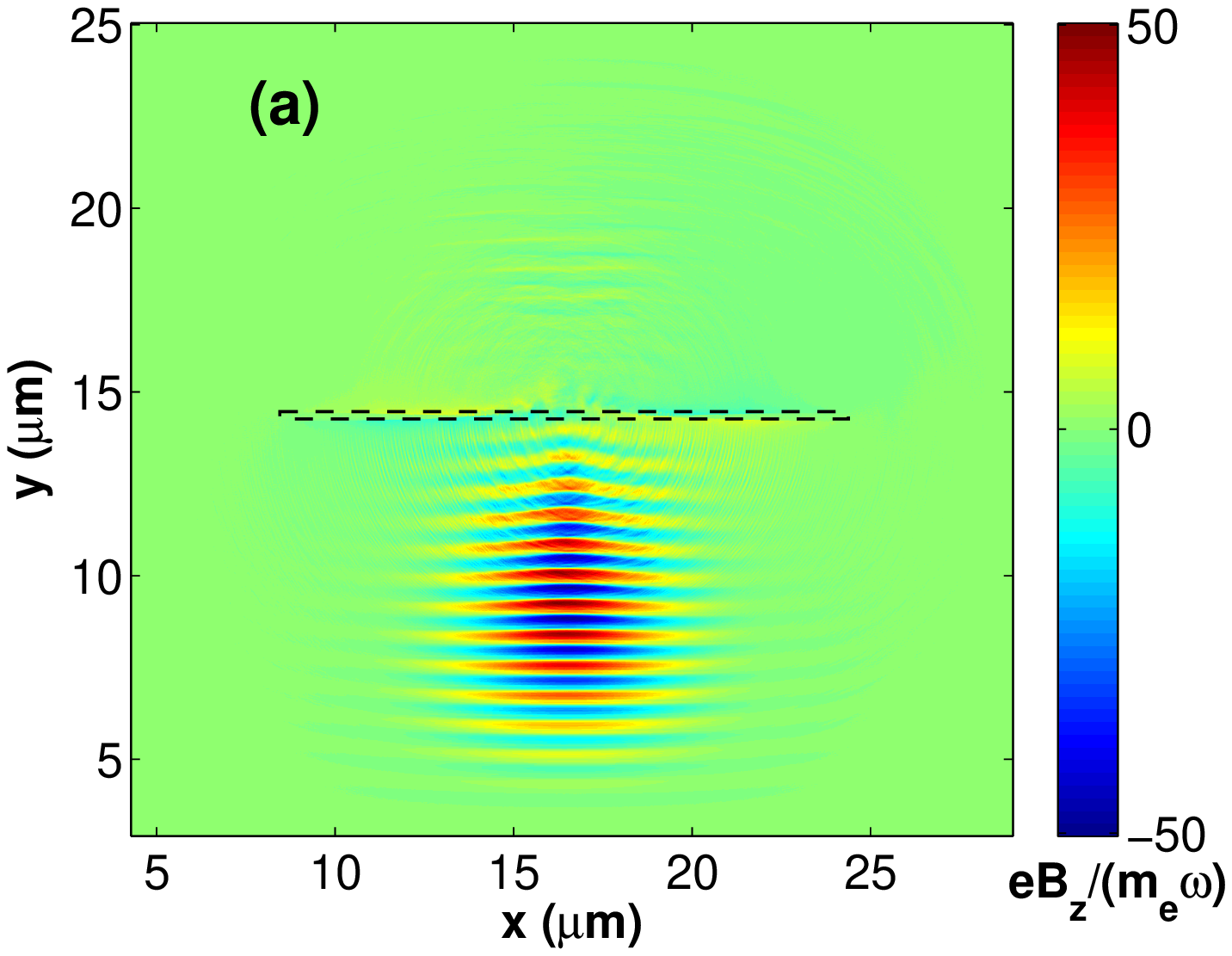}
\includegraphics[width=0.45\textwidth]{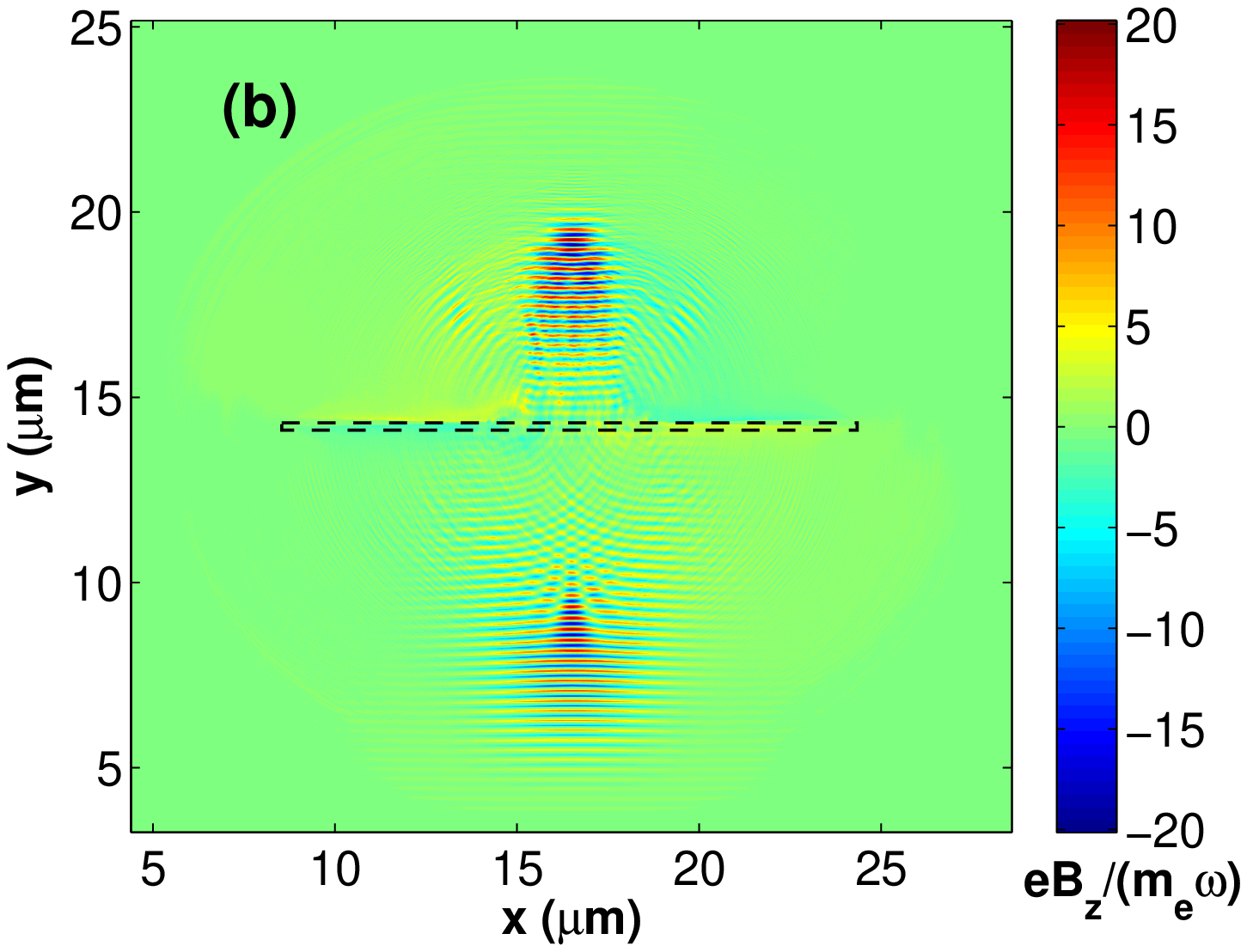}
\caption{\label{fig1} (Color online) Magnetic field $B_z$ in the simulation box showing reflected and transmitted part of the laser pulse after its interaction with $200~\rm{nm}$ thick foil of electron density $3.5 \cdot 10^{23}~\rm{cm^{-3}}$ for (a) $1 \omega$ and (b) $3 \omega$ cases. 
Initial position of the foil is marked by the dashed line.}
\end{center}
\end{figure}

\begin{figure}
\begin{center}
\includegraphics[width=0.45\textwidth]{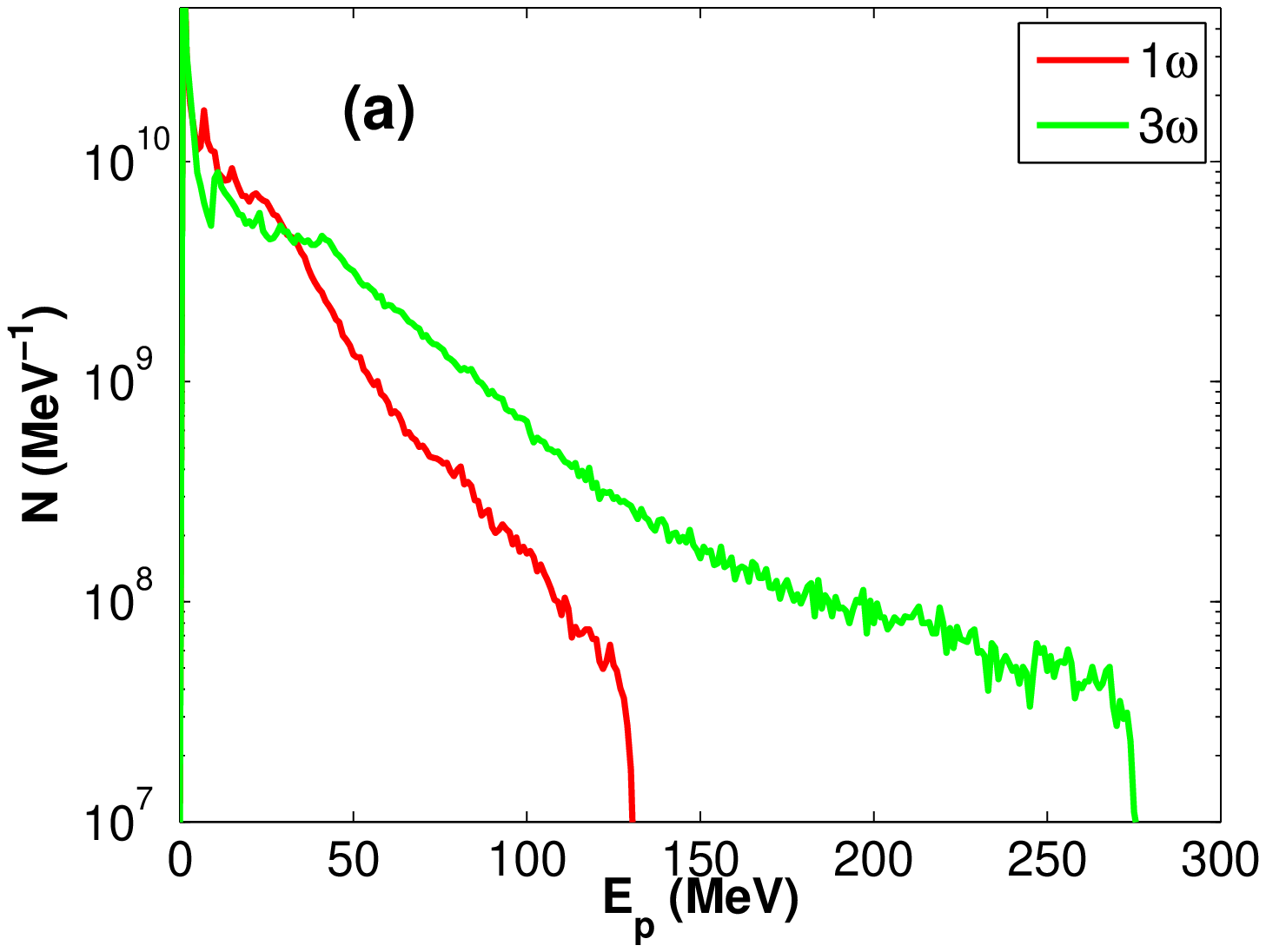}
\includegraphics[width=0.45\textwidth]{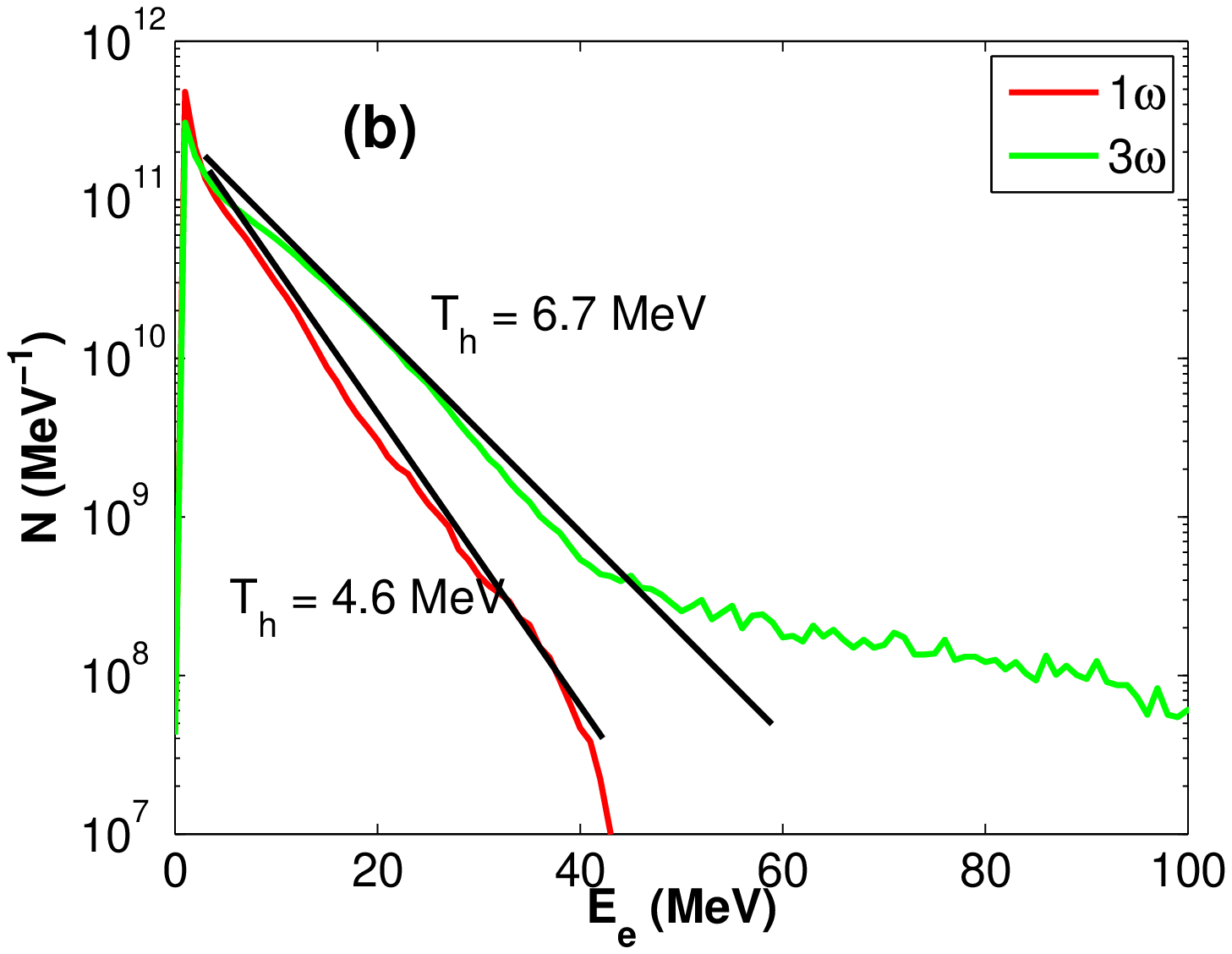}
\caption{\label{fig2} (Color online) Energy spectra of (a) accelerated protons in the forward direction at the end of simulations (approx. $100~\rm{fs}$ after laser-foil interaction), and (b) heated electrons at the target rear (non-irradiated) side at the end of laser-target interaction for $1 \omega$ and $3 \omega$ cases and target thickness $200~\rm{nm}$. 
Hot electrons can be described by temperature $T_h$ except for high-energy tail at $3 \omega$ case. }
\end{center}
\end{figure}

Firstly, we illustrate the difference between using $1 \omega$ and $3 \omega$ pulses for the foil of thickness $200~\rm{nm}$. 
At $1 \omega$ ($\lambda_1=800~\rm{nm}$), most of the laser pulse is reflected from the target, but a small part of the pulse can be transmitted through (amplitude $a_0 \approx 5$). 
At $3 \omega$ ($\lambda_3=264~\rm{nm}$), large part of the laser pulse is transmitted through the target (the foil is "punched through" before the end of the laser-target interaction) due to the induced transparency, see Fig. \ref{fig1}. 
Although $a_0$ is reduced almost three times by switching the laser frequency, the maximum proton energy in the forward direction is increased from $132~\rm{MeV}$ at $1 \omega$ case to $277~\rm{MeV}$ at $3 \omega$ case (see proton energy spectra in Fig. \ref{fig2}a). 
This strongly differs from the standard model of Target Normal Sheath Acceleration (TNSA) mechanism \cite{Snavely2000}, where the maximum ion energy is directly proportional to $\sim a_0$ by assuming ponderomotive scaling of hot electron temperature \cite{Wilks2001}. 
Moreover, the number of accelerated ions can be also substantially increased. 
For example, the number of protons with energy higher than $60~\rm{MeV}$ already suitable for proton therapy applications is enhanced by factor 4 in $3 \omega$ compared to $1 \omega$ case for the used target thickness. 
Thus, two different ion acceleration regimes take place for the same foil thickness and laser pulse intensity, but different laser frequency. 
For $3 \omega$ pulse, an enhanced ion acceleration connected with relativistic transparency, such as Laser Breakout Afterburner (BOA), should be operative \cite{Jung2013b}.

The enhancement of proton energy and number can be explained by a more efficient electron heating in $3 \omega$ case due to relativistically induced transparency. 
When the target is relativistically transparent, electrons can be accelerated by the ponderomotive force in the whole target volume (given by the target thickness), whereas they are accelerated only in the surface layer determined by the skin depth for opaque targets. 
Therefore, the work done by the ponderomotive force on the electrons is higher for thicker relativistically transparent targets ($3 \omega$ case) even if $a_0$ is smaller than for the opaque target ($1 \omega$ case).
In the energy spectra of electrons which are located behind the rear side of the foil (Fig. \ref{fig2}b), two distinct populations of hot electrons can be identified for $3 \omega$ whereas only one population appears for $1 \omega$ case. 
The first population can be approximated by $N_e \approx \exp{(-E_k/T_h)}$, where the hot electron temperature $T_h$ is about $4.6~\rm{MeV}$ and $6.7~\rm{MeV}$ in $1 \omega$ and $3 \omega$ case, respectively. 
Thus, one can observe that the temperature and number of hot electrons in the first population is enhanced for $3 \omega$ (relativistically transparent) compared with $1 \omega$ (nontransparent) case, which also corresponds to the increased absorption of laser pulse energy in the plasma (45\% vs. 21\% for $3 \omega$ and $1 \omega$, respectively). 
The high-energy tail in the spectrum for $3 \omega$ case, which can be regarded as the second population of hot electrons, corresponds to the electron bunches directly accelerated by the propagating laser wave beyond the target similarly to the case of the so-called direct laser acceleration \cite{Pukhov1999}. 

We should note that this mechanism works also when a small target preplasma at the front or rear side was assumed in the simulations instead of initial step-like density plasma profile. 
For example, when an exponential density profile $\exp{(-x/L)}$ with the scale length $L = 80~\rm{nm}$ was initialized and the thickness of the layer with constant maximum density was substantially reduced in order to keep the areal density of the target the same in the simulations, the maximum energy of accelerated protons only slightly differs (about 5\% at most) for any combination of step-like and exponential density profiles on the target front/rear sides.

\subsection{Dependence on the target thickness}
\begin{figure}
\begin{center}
\includegraphics[width=0.45\textwidth]{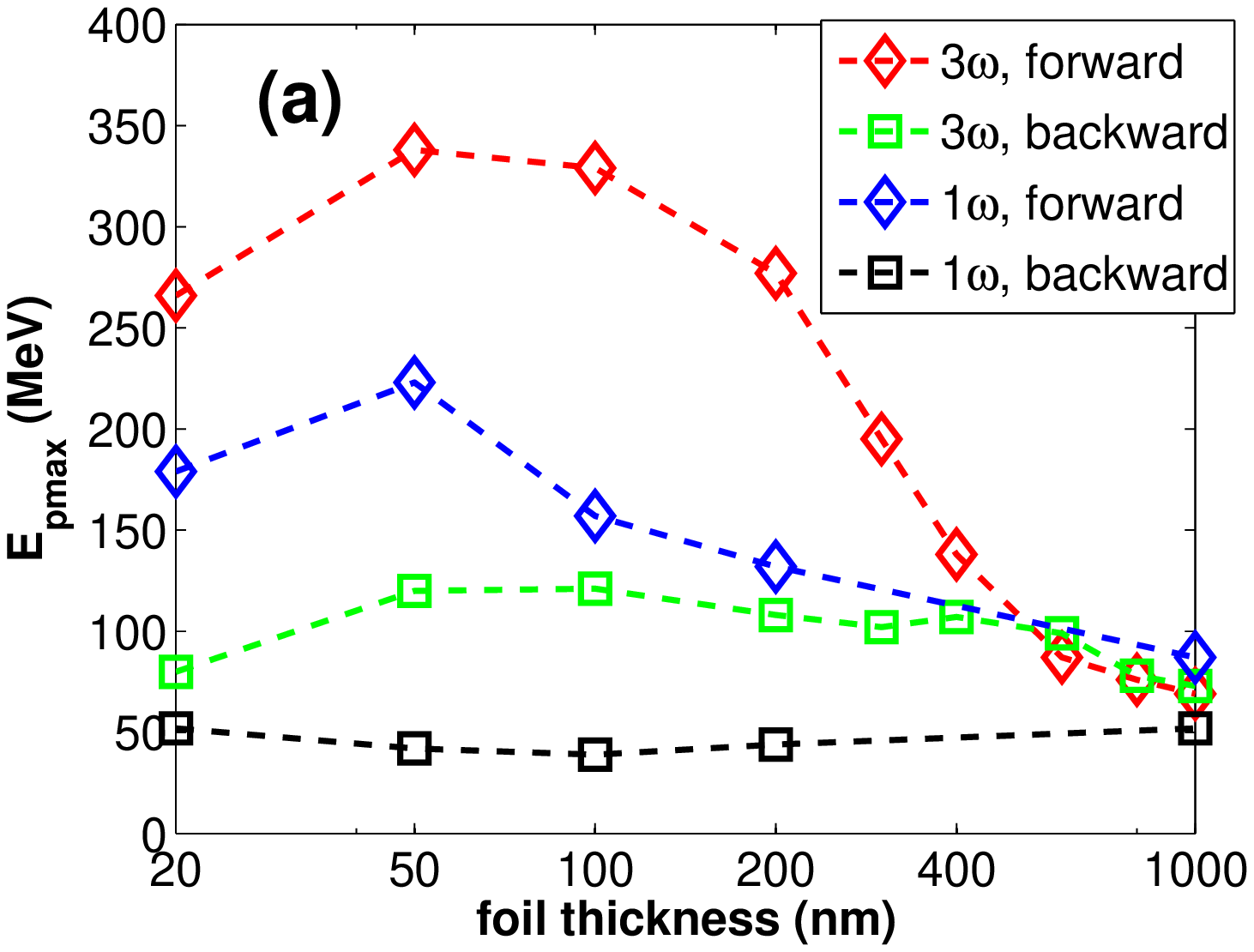}
\includegraphics[width=0.45\textwidth]{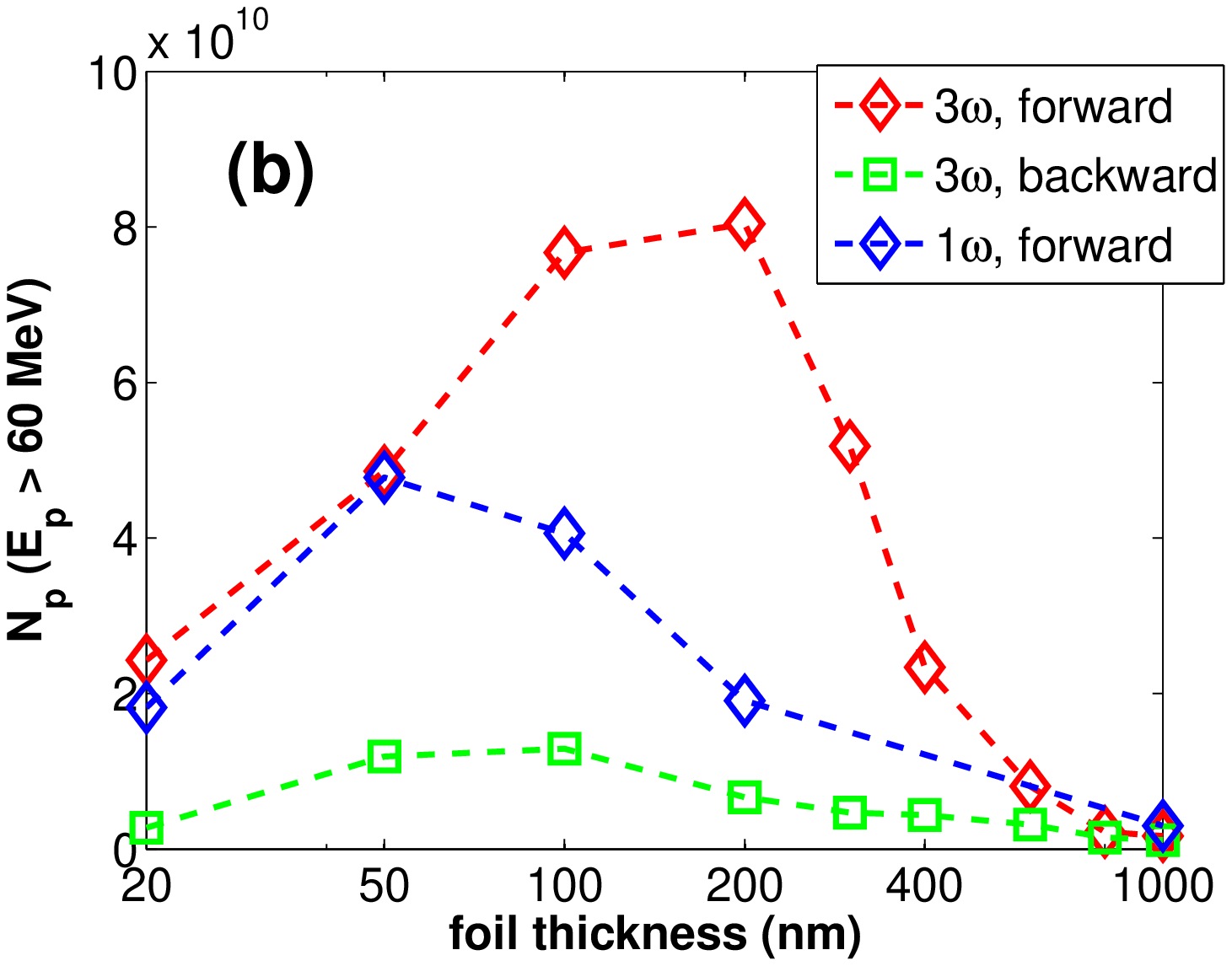}
\includegraphics[width=0.45\textwidth]{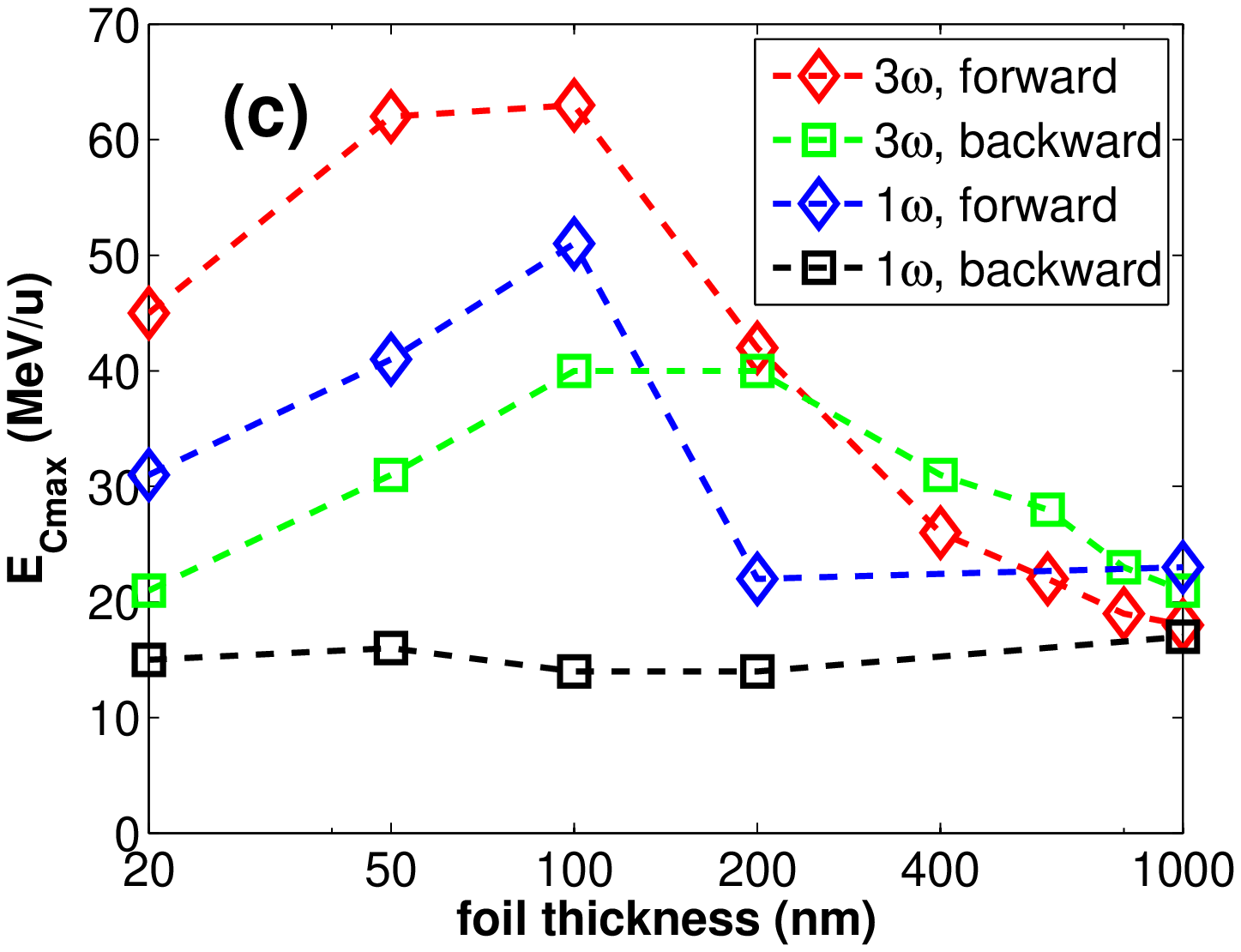}
\includegraphics[width=0.45\textwidth]{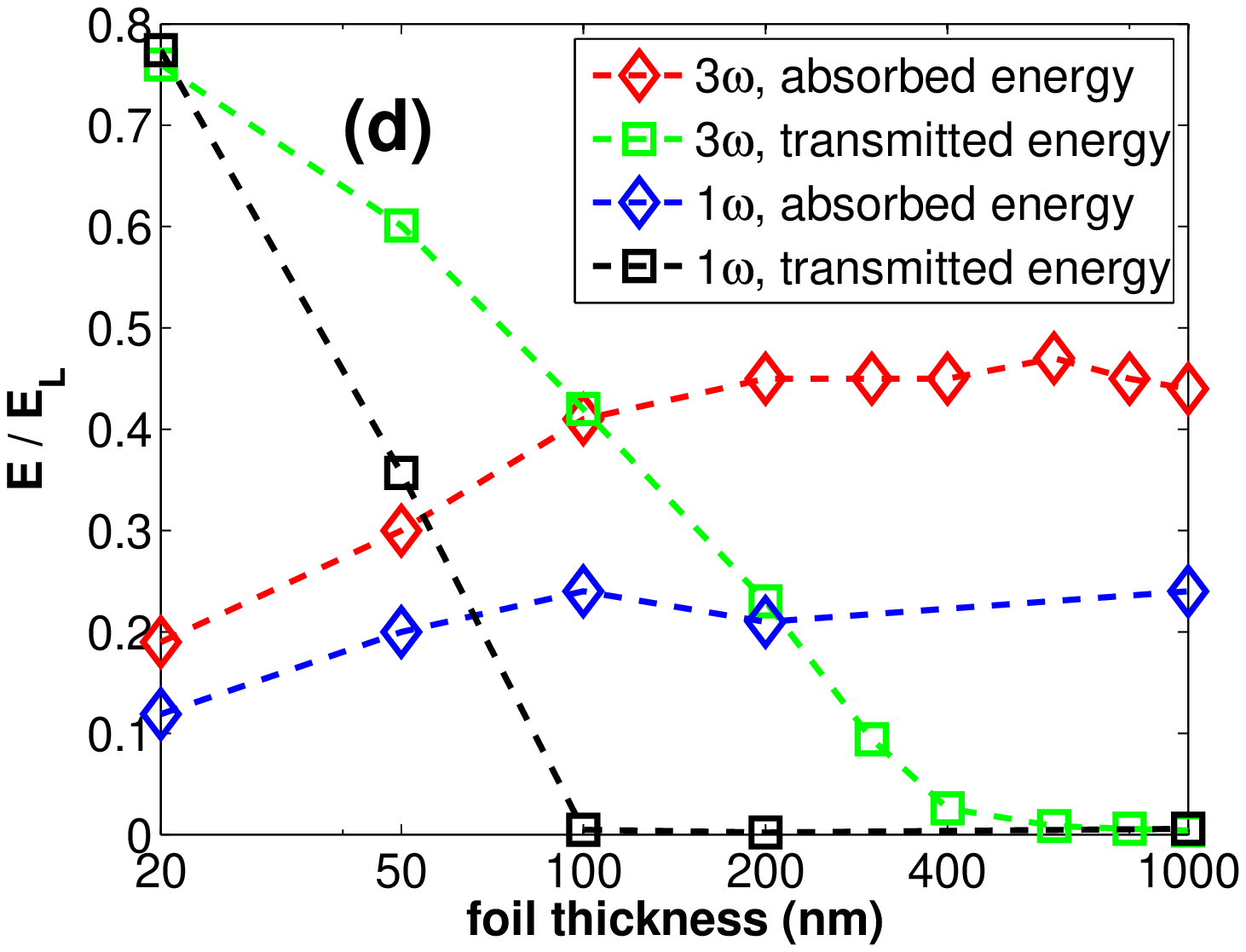}
\caption{\label{fig3} (Color online) Dependence of (a) maximum proton energy and (b) numbers of high energy protons ($\rm{energy} > 60~\rm{MeV}$), (c) maximum energy of $\rm{C^{6+}}$ ions accelerated in the forward / backward direction on the thickness of ionized plastic foil ($\rm{CH_2}$) irradiated by $1 \omega$ and $3 \omega$ laser pulse. 
(d) Dependence of the laser pulse energy transmitted through the foil and the absorbed energy in plasma on the foil thickness. 
The following thickness of the foil can be used in the simulations: $20~\rm{nm}$, $50~\rm{nm}$, $100~\rm{nm}$, $200~\rm{nm}$, $300~\rm{nm}$, $400~\rm{nm}$, $600~\rm{nm}$, $800~\rm{nm}$, $1~\rm{\mu m}$.}
\end{center}
\end{figure}

The efficiency of ion acceleration varies with the target thickness as shown in several experimental and theoretical studies \cite{Neely2006, Ceccotti2007}. 
Moreover, one may obtain induced transparency with thinner targets more easily \cite{Yin2006}, especially when the target density is slightly above the threshold for induced transparency as in our case. 
Therefore, in the following set of simulations, we illustrate the difference between using $1 \omega$ and $3 \omega$ pulses for various thicknesses of plastic foils ranging from $20~\rm{nm}$ to $1~\rm{\mu m}$. 
Fig. \ref{fig3} shows the dependence of (a) maximum energies, (b) numbers of high-energy protons, and (c) maximum energies of $\rm{C^{6+}}$ ions accelerated in the forward and backward directions (with respect to the propagation direction of the incident laser beam) as well as (d) the dependence of the ratio of the absorbed and transmitted laser pulse energy to the total laser beam energy on the foil thickness and laser frequency. 
The simulations show that a noticeable part of the laser pulse can propagate through the foil up to the thickness of $400~\rm{nm}$ for $3 \omega$, whereas the thickness less than $100~\rm{nm}$ is required for $1 \omega$ case. 
Note that the plasma skin depth $c / \omega_{pe}$ \cite{Gibbon2005}, where $\omega_{pe} \approx \sqrt{e^2n_e/(\epsilon_0\gamma_L m_e)}$ assuming relativistic mass of electrons oscillating in the linearly polarized laser field with relativistic factor $\gamma_L \approx \sqrt{1+a_0^2/2}$, is about $60~\rm{nm}$ and $35~\rm{nm}$ for $1 \omega$ and $3 \omega$ case, respectively. 
Thus, the transmission of a substantial part of laser pulse energy through the foil could be explained by the thickness of the foil smaller than the skin depth for $1 \omega$ case. 
On the contrary, only relativistically induced transparency is able to explain this effect for the foils thicker than $50~\rm{nm}$ in the $3 \omega$ case. 

For laser frequency $3 \omega$, the enhancement of maximum energy and the number of accelerated high-energy protons roughly correlates with the amount of transmitted laser pulse energy and with the asymmetry of ion acceleration from the front and rear sides of the fully ionized polyethylene foils (in the backward and forward directions, respectively). 
Only 3\% of the laser pulse energy incident on the target is transmitted through the foil in the case of $400~\rm{nm}$ thickness. 
In fact, only a small rear part of the pulse propagates through with maximum amplitude $a_0 \approx 10$. 
The amount of transmitted laser pulse energy as well as the length of the transmitted part of the pulse (and its amplitude) significantly increases with decreasing thickness. 
For $300~\rm{nm}$ thick foil, the amount of transmitted laser pulse energy reaches almost 10\% of the incident energy and the maximum amplitude of the transmitted part of the pulse is $a_0 \approx 15$. 
For $50~\rm{nm}$ foil, 60\% of the pulse energy is transmitted and the maximum transmitted pulse amplitude is close to the amplitude of the incident laser pulse. 
However, the amount of the absorbed energy in the plasma is reduced for the thinnest foils as the density of the expanding foil decreases rapidly already during laser-target interaction. 
Reduced laser pulse absorption and relatively low number of particles in the laser focal spot leads to the decrease of the number of high-energy protons even if the maximum energy only slightly falls for the thinnest foils.

At fundamental laser frequency, the enhancement of proton energy and number can be also observed with decreasing target thickness. 
However, such enhancement is less pronounced than for the third harmonic frequency. 
The increase in maximum energy for the foil of thickness $200~\rm{nm}$ compared with $1~\rm{\mu m}$ about 50\% in the forward direction is in qualitative agreement with previous experimental observations at significantly lower maximum intensity of the pulse ($\sim 10^{19}~\rm{W/cm^2}$) \cite{Neely2006}. 
The most efficient proton acceleration takes place at the foil thickness $50~\rm{nm}$ when the foil becomes transparent already during laser-target interaction due to its rapid expansion. 
In $3 \omega$ case, the highest proton energy can be also observed for $50~\rm{nm}$ foil. 
Nevertheless, the optimum thickness for $3 \omega$ is between $100~\rm{nm}$ and $200~\rm{nm}$, since there is a clear maximum of the number of high-energy protons whereas the maximum energy is only slightly lower compared with $50~\rm{nm}$ foil. 

When a part of the laser pulse is transmitted through the foil, one can observe a strong asymmetry in terms of maximum energy of accelerated ions in the forward/backward direction (Fig. \ref{fig3}a). 
We observed this asymmetry when the transmitted part has amplitude $a_0 \gg 1$ (when the $v \times B$ term in the Lorentz force becomes significant as the electric term). 
At amplitude $a_0 \gg 1$, electrons oscillating in the laser wave have relativistic quiver velocity \cite{Gibbon2005}, their trajectories can be bent by the magnetic field of the wave towards laser propagation direction and their energy is enhanced. 
Such increase of electron energy is translated into fast ions. 
The transmitted laser wave with amplitude $a_0 \gg 1$ is observed in our simulations only for targets with thickness equal or less than $400~\rm{nm}$ at $3 \omega$ case, whereas it is observed for all studied foils at $1 \omega$ (even if the amount of transmitted laser pulse energy is below 1\% for the foils thicker than $50~\rm{nm}$). 

The dependence of maximum energy on the foil thickness is quite similar for heavier $\rm{C^{6+}}$ ions (see Fig. \ref{fig3}c). 
Since the protons from target surface layer are accelerated prior to heavier ions due to their higher charge to mass ratio, the protons partly shield accelerating electric field for $\rm{C^{6+}}$ ions. 
Therefore, the maximum energy of $\rm{C^{6+}}$ ions per atomic mass unit is reduced on 20\%-40\% of the energy of protons. 
One can also observe that the ratio of the energy of $\rm{C^{6+}}$ ions accelerated in the backward direction to the energy of the ions accelerated in the forward direction is slightly enhanced compared with protons. 
The difference can be ascribed to acceleration of protons prior carbon ions towards target interior from the front (laser-irradiated) side due to radiation pressure in the initial stage of interaction, which subsequently leads to a relatively small amount of light ions (protons) on the target front surface compared with the amount of heavier ions. 
After laser-target interaction, a significantly lower number of protons on the front target side shields the accelerating electric field on carbon ions less than on the rear target side.

\subsection{Dependence on the target density at $3 \omega$ case}
In the simulations described above, the target density was fixed at $n_e=3.5 \cdot 10^{23}~\rm{cm^{-3}}$, which is a realistic density for fully ionized polyethylene $\rm{CH_2}$ foil. 
However, this density was slightly above theoretical threshold for the induced transparency for a given laser pulse amplitude. 
Therefore, in the second set of simulations, we investigated the using of the third harmonic frequency for various target densities (from $4.375 \cdot 10^{22}~\rm{cm^{-3}}$ to $7.0 \cdot 10^{23}~\rm{cm^{-3}}$) with fixed thickness of the foil equal to $200~\rm{nm}$. 
Other parameters have been kept the same as in previous simulations. 

\begin{figure}
\begin{center}
\includegraphics[width=0.45\textwidth]{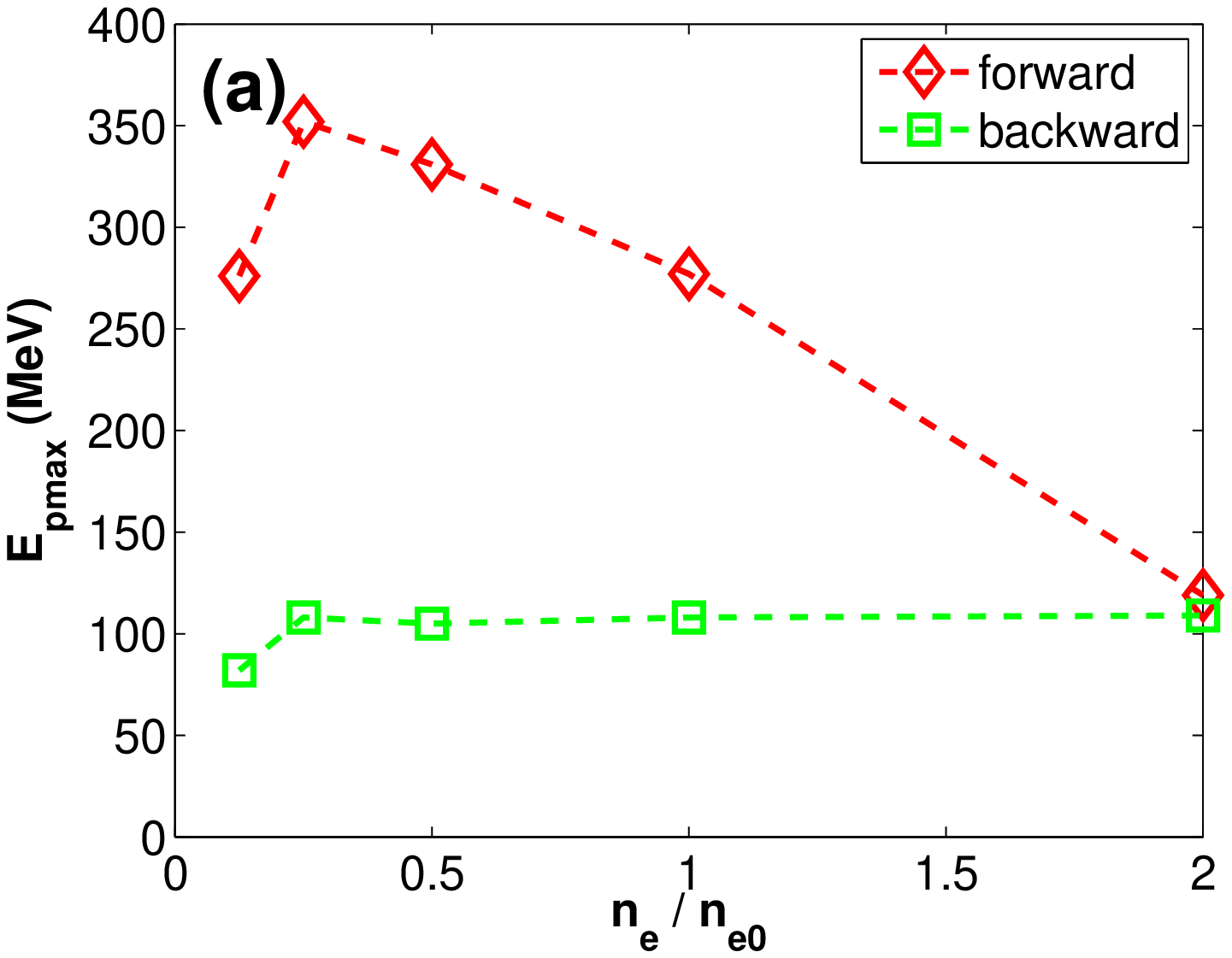}
\includegraphics[width=0.45\textwidth]{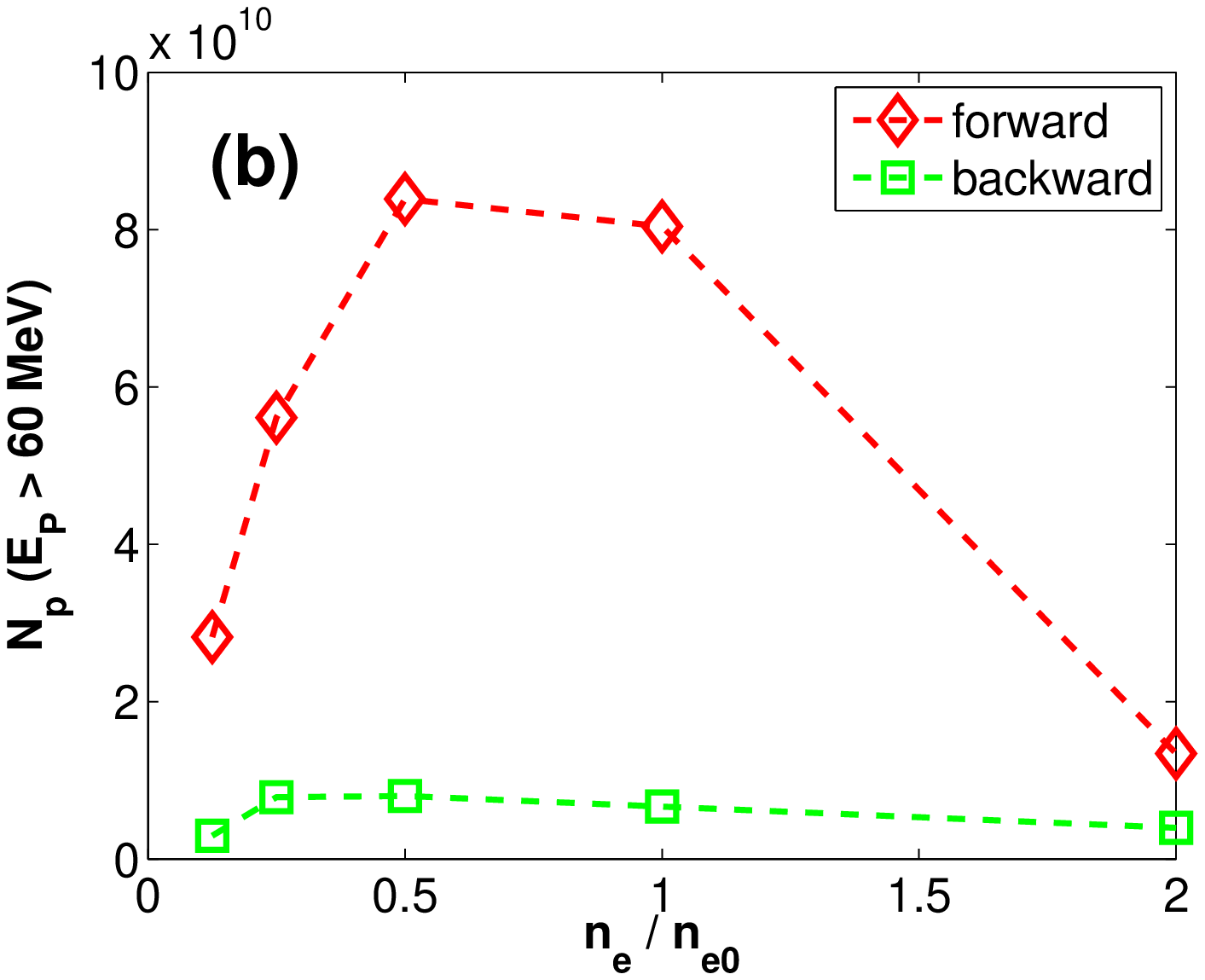}
\includegraphics[width=0.45\textwidth]{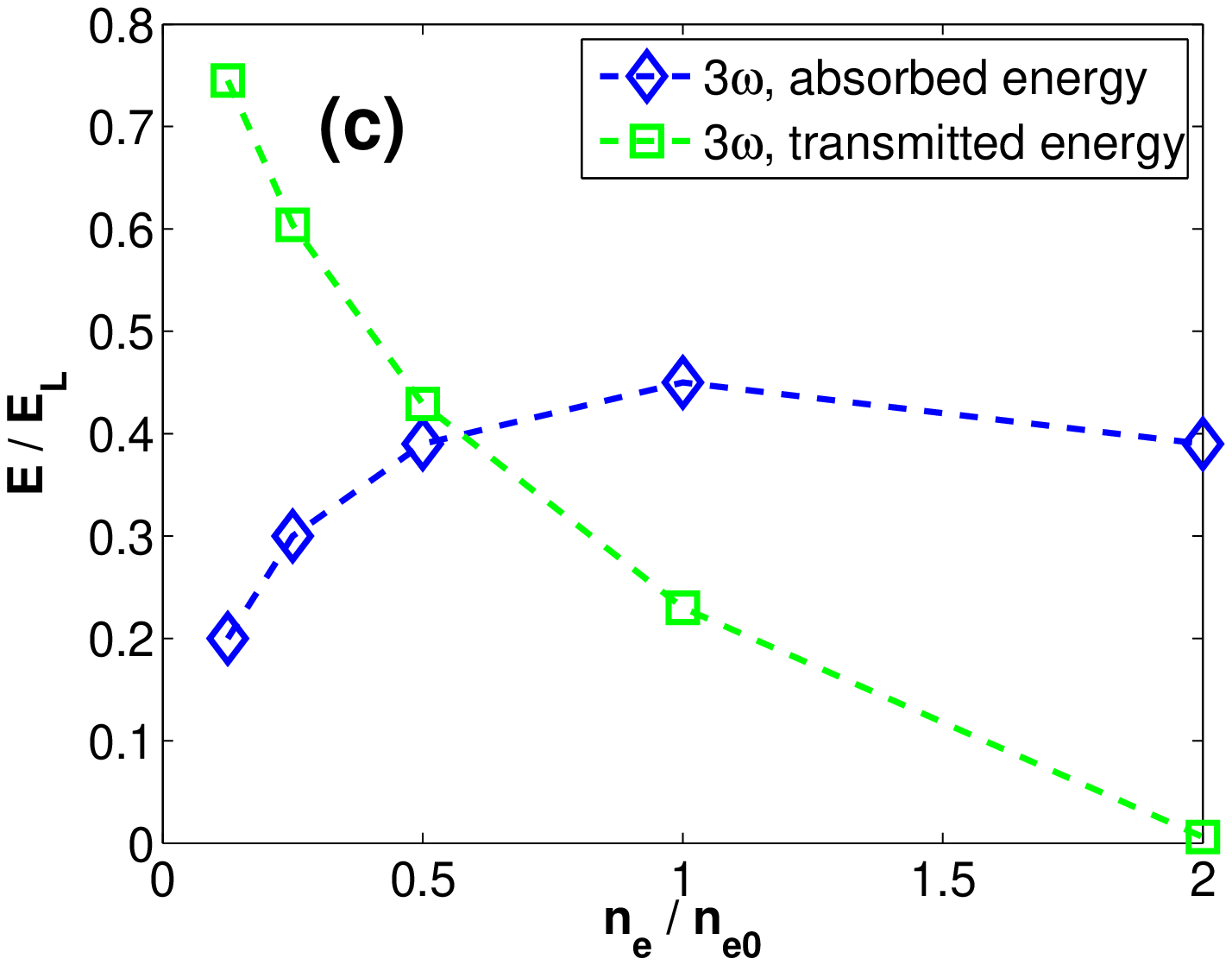}
\caption{\label{fig4} (Color online) Dependence of (a) maximum proton energy and (b) numbers of high energy protons ($\rm{energy} > 60~\rm{MeV}$) accelerated in the forward / backward direction, (c) laser pulse energy transmitted through the foil and the absorbed energy in plasma on the density of $200~\rm{nm}$ thick ionized plastic foil ($\rm{CH_2}$) irradiated by $3 \omega$ laser pulse. 
Initial target density is normalized to the electron density $n_{e0} = 3.5 \cdot 10^{23}~\rm{cm^{-3}}$, which is the density used in previous set of simulations. }
\end{center}
\end{figure}

One can observe similarly to previous set of simulations that the ion acceleration is enhanced in the forward direction when a part of the pulse is transmitted and, simultaneously, when it has a larger amplitude ($a_0 \gg 1$) after propagation through the target. 
For electron density $n_e=3.5 \cdot 10^{23}~\rm{cm^{-3}}$, the maximum energy of accelerated protons is equal to $277~\rm{MeV}$, whereas it is only $108~\rm{MeV}$ in the backward direction (see Fig. \ref{fig4}a). 
When we increase the target density by factor 2 up to $n_e=7.0 \cdot 10^{23}~\rm{cm^{-3}}$, the laser pulse is not transmitted substantially through the target (electromagnetic wave with amplitude $a_0 \approx 1$ propagates behind the target) and the maximum energy of protons accelerated in the forward and backward directions is almost the same $-~119~\rm{MeV}$ and $109~\rm{MeV}$, respectively (but the number of accelerated protons is substantially higher in the forward direction). 
On the contrary, when the density of the foil is twice or even four times decreased, proton energy in the forward direction increases further about 20\% or 30\%, respectively. 
Then, the energy starts to decrease with the density. 
In the backward direction, the energy is constant except for the lowest density case $n_e=4.375 \cdot 10^{22}~\rm{cm^{-3}}$, where it decreases about 20\%. 
Thus, we found the maximum energy of accelerated protons for electron density of the foil $n_e=8.75 \cdot 10^{22}~\rm{cm^{-3}}$ ($\approx 5.4~n_c$ for $3 \omega$). 
However, for another important parameter - the number of high energy protons (with energy higher than $60~\rm{MeV}$ suitable for proton therapy), optimal target density between $1.75 \cdot 10^{23}$ and $3.5 \cdot 10^{23}~\rm{cm^{-3}}$ ($\approx 10.9~n_c$ and $21.8~n_c$ for $3 \omega$, respectively) was found in our simulations as can be seen in Fig. \ref{fig4}b. 
It corresponds to the highest absorption of laser pulse energy into the plasma (39\% and 45\%, respectively). 
We should note that it does not mean optimal values for all laser and target parameters. 
The investigation of such dependency on various laser and target parameters is beyond the scope of this paper. 
For example, one may suppose that the using of the 4th harmonics may be more beneficial when the target of higher density (e.g., metal foils) would be assumed.

\section{Conclusion}
In conclusion, we have proposed a significant enhancement of ion acceleration from ionized solid target by using ultrashort intense (subPW) laser pulse at higher harmonic frequency. 
In our numerical simulations, we have demonstrated the increase of maximum proton energy as well as the increase of the number of high-energy protons suitable for proton therapy by factor 2 by using the third harmonic frequency compared with fundamental frequency at optimal thickness of ionized plastic foil. 
Higher energies and numbers of accelerated ions are explained by a more efficient electron heating and acceleration by the ponderomotive force in the whole target volume (given by the target thickness) when the target is relativistically transparent. 
Since the laser intensity required for relativistically induced transparency is directly proportional to $\lambda^2 \sim 1/\omega^2$ (where $\lambda$ is the laser wavelength and $\omega$ is the frequency), this regime is achievable more easily with higher (harmonic) laser frequency than the fundamental one.

Such strong enhancement of ion acceleration can surpass other enhancements which propose to use special targets like the ones with microstructures \cite{Klimo2011, Margarone2012, Floquet2013}, foam \cite{Sgattoni2012}, or grating \cite{Ceccotti2013} on the surface. 
Moreover, flat foil targets can be more easily produced than other special targets, which is important from the point of view of future applications of high repetition rate femtosecond lasers.


%
%

%

\begin{acknowledgments}
This research has been supported by the Czech Science Foundation (project No. P205/12/P366) and by the Ministry of Education, Youth and Sports of the Czech Republic under projects ELI-Beamlines (CZ.1.05/1.1.00/02.0061), ECOP (CZ.1.07/2.3.00/20.0087). 
The authors would like to acknowledge IT4Innovations Centre of Excellence project (CZ.1.05/1.1.00/02.0070) funded by the European Regional Development Fund as well as Large Research, Development and Innovation Infrastructures project (LM2011033) funded by the Ministry of Education, Youth and Sports of the Czech Republic for providing computing resources. 
Also, access to computing and storage facilities owned by parties and projects contributing to the National Grid Infrastructure MetaCentrum, provided under the programme Projects of Large Infrastructure for Research, Development, and Innovations (LM2010005), is greatly appreciated.
\end{acknowledgments}


\begin{thebibliography}{37}%
\makeatletter
\providecommand \@ifxundefined [1]{%
 \@ifx{#1\undefined}
}%
\providecommand \@ifnum [1]{%
 \ifnum #1\expandafter \@firstoftwo
 \else \expandafter \@secondoftwo
 \fi
}%
\providecommand \@ifx [1]{%
 \ifx #1\expandafter \@firstoftwo
 \else \expandafter \@secondoftwo
 \fi
}%
\providecommand \natexlab [1]{#1}%
\providecommand \enquote  [1]{``#1''}%
\providecommand \bibnamefont  [1]{#1}%
\providecommand \bibfnamefont [1]{#1}%
\providecommand \citenamefont [1]{#1}%
\providecommand \href@noop [0]{\@secondoftwo}%
\providecommand \href [0]{\begingroup \@sanitize@url \@href}%
\providecommand \@href[1]{\@@startlink{#1}\@@href}%
\providecommand \@@href[1]{\endgroup#1\@@endlink}%
\providecommand \@sanitize@url [0]{\catcode `\\12\catcode `\$12\catcode
  `\&12\catcode `\#12\catcode `\^12\catcode `\_12\catcode `\%12\relax}%
\providecommand \@@startlink[1]{}%
\providecommand \@@endlink[0]{}%
\providecommand \url  [0]{\begingroup\@sanitize@url \@url }%
\providecommand \@url [1]{\endgroup\@href {#1}{\urlprefix }}%
\providecommand \urlprefix  [0]{URL }%
\providecommand \Eprint [0]{\href }%
\providecommand \doibase [0]{http://dx.doi.org/}%
\providecommand \selectlanguage [0]{\@gobble}%
\providecommand \bibinfo  [0]{\@secondoftwo}%
\providecommand \bibfield  [0]{\@secondoftwo}%
\providecommand \translation [1]{[#1]}%
\providecommand \BibitemOpen [0]{}%
\providecommand \bibitemStop [0]{}%
\providecommand \bibitemNoStop [0]{.\EOS\space}%
\providecommand \EOS [0]{\spacefactor3000\relax}%
\providecommand \BibitemShut  [1]{\csname bibitem#1\endcsname}%
\let\auto@bib@innerbib\@empty
\bibitem [{\citenamefont {Macchi}\ \emph {et~al.}(2013)\citenamefont {Macchi},
  \citenamefont {Borghesi},\ and\ \citenamefont {Passoni}}]{Macchi2013}%
  \BibitemOpen
  \bibfield  {author} {\bibinfo {author} {\bibfnamefont {A.}~\bibnamefont
  {Macchi}}, \bibinfo {author} {\bibfnamefont {M.}~\bibnamefont {Borghesi}}, \
  and\ \bibinfo {author} {\bibfnamefont {M.}~\bibnamefont {Passoni}},\ }\href
  {\doibase 10.1103/RevModPhys.85.751} {\bibfield  {journal} {\bibinfo
  {journal} {Rev. Mod. Phys.}\ }\textbf {\bibinfo {volume} {85}},\ \bibinfo
  {pages} {751} (\bibinfo {year} {2013})}\BibitemShut {NoStop}%
\bibitem [{\citenamefont {Cirrone}\ \emph {et~al.}(2013)\citenamefont
  {Cirrone}, \citenamefont {Carpinelli}, \citenamefont {Cuttone}, \citenamefont
  {Gammino}, \citenamefont {Jia}, \citenamefont {Korn}, \citenamefont
  {Maggiore}, \citenamefont {Manti}, \citenamefont {Margarone}, \citenamefont
  {Prokupek}, \citenamefont {Renis}, \citenamefont {Romano}, \citenamefont
  {Schillaci}, \citenamefont {Tomasello}, \citenamefont {Torrisi},
  \citenamefont {Tramontana},\ and\ \citenamefont {Velyhan}}]{Cirrone2013}%
  \BibitemOpen
  \bibfield  {author} {\bibinfo {author} {\bibfnamefont {G.~A.~P.}\
  \bibnamefont {Cirrone}}, \bibinfo {author} {\bibfnamefont {M.}~\bibnamefont
  {Carpinelli}}, \bibinfo {author} {\bibfnamefont {G.}~\bibnamefont {Cuttone}},
  \bibinfo {author} {\bibfnamefont {S.}~\bibnamefont {Gammino}}, \bibinfo
  {author} {\bibfnamefont {S.~B.}\ \bibnamefont {Jia}}, \bibinfo {author}
  {\bibfnamefont {G.}~\bibnamefont {Korn}}, \bibinfo {author} {\bibfnamefont
  {M.}~\bibnamefont {Maggiore}}, \bibinfo {author} {\bibfnamefont
  {L.}~\bibnamefont {Manti}}, \bibinfo {author} {\bibfnamefont
  {D.}~\bibnamefont {Margarone}}, \bibinfo {author} {\bibfnamefont
  {J.}~\bibnamefont {Prokupek}}, \bibinfo {author} {\bibfnamefont
  {M.}~\bibnamefont {Renis}}, \bibinfo {author} {\bibfnamefont
  {F.}~\bibnamefont {Romano}}, \bibinfo {author} {\bibfnamefont
  {F.}~\bibnamefont {Schillaci}}, \bibinfo {author} {\bibfnamefont
  {B.}~\bibnamefont {Tomasello}}, \bibinfo {author} {\bibfnamefont
  {L.}~\bibnamefont {Torrisi}}, \bibinfo {author} {\bibfnamefont
  {A.}~\bibnamefont {Tramontana}}, \ and\ \bibinfo {author} {\bibfnamefont
  {A.}~\bibnamefont {Velyhan}},\ }\href {\doibase 10.1016/j.nima.2013.05.051}
  {\bibfield  {journal} {\bibinfo  {journal} {Nucl. Ins. Meth. Phys. Res. A}\
  }\textbf {\bibinfo {volume} {730}},\ \bibinfo {pages} {174} (\bibinfo {year}
  {2013})}\BibitemShut {NoStop}%
\bibitem [{\citenamefont {Ledingham}\ and\ \citenamefont
  {Galster}(2010)}]{Ledingham2010}%
  \BibitemOpen
  \bibfield  {author} {\bibinfo {author} {\bibfnamefont {K.~W.~D.}\
  \bibnamefont {Ledingham}}\ and\ \bibinfo {author} {\bibfnamefont
  {W.}~\bibnamefont {Galster}},\ }\href {\doibase
  10.1088/1367-2630/12/4/045005} {\bibfield  {journal} {\bibinfo  {journal}
  {New J. Phys.}\ }\textbf {\bibinfo {volume} {12}},\ \bibinfo {pages} {045005}
  (\bibinfo {year} {2010})}\BibitemShut {NoStop}%
\bibitem [{\citenamefont {Snavely}\ \emph {et~al.}(2007)\citenamefont
  {Snavely}, \citenamefont {Zhang}, \citenamefont {Akli}, \citenamefont {Chen},
  \citenamefont {Freeman}, \citenamefont {Gu}, \citenamefont {Hatchett},
  \citenamefont {Hey}, \citenamefont {Hill}, \citenamefont {Key}, \citenamefont
  {Izawa}, \citenamefont {King}, \citenamefont {Kitagawa}, \citenamefont
  {Kodama}, \citenamefont {Langdon}, \citenamefont {Lasinski}, \citenamefont
  {Lei}, \citenamefont {MacKinnon}, \citenamefont {Patel}, \citenamefont
  {Stephens}, \citenamefont {Tampo}, \citenamefont {Tanaka}, \citenamefont
  {Town}, \citenamefont {Toyama}, \citenamefont {Tsutsumi}, \citenamefont
  {Wilks}, \citenamefont {Yabuuchi},\ and\ \citenamefont
  {Zheng}}]{Snavely2007}%
  \BibitemOpen
  \bibfield  {author} {\bibinfo {author} {\bibfnamefont {R.~A.}\ \bibnamefont
  {Snavely}}, \bibinfo {author} {\bibfnamefont {B.}~\bibnamefont {Zhang}},
  \bibinfo {author} {\bibfnamefont {K.}~\bibnamefont {Akli}}, \bibinfo {author}
  {\bibfnamefont {Z.}~\bibnamefont {Chen}}, \bibinfo {author} {\bibfnamefont
  {R.~R.}\ \bibnamefont {Freeman}}, \bibinfo {author} {\bibfnamefont
  {P.}~\bibnamefont {Gu}}, \bibinfo {author} {\bibfnamefont {S.~P.}\
  \bibnamefont {Hatchett}}, \bibinfo {author} {\bibfnamefont {D.}~\bibnamefont
  {Hey}}, \bibinfo {author} {\bibfnamefont {J.}~\bibnamefont {Hill}}, \bibinfo
  {author} {\bibfnamefont {M.~H.}\ \bibnamefont {Key}}, \bibinfo {author}
  {\bibfnamefont {Y.}~\bibnamefont {Izawa}}, \bibinfo {author} {\bibfnamefont
  {J.}~\bibnamefont {King}}, \bibinfo {author} {\bibfnamefont {Y.}~\bibnamefont
  {Kitagawa}}, \bibinfo {author} {\bibfnamefont {R.}~\bibnamefont {Kodama}},
  \bibinfo {author} {\bibfnamefont {A.~B.}\ \bibnamefont {Langdon}}, \bibinfo
  {author} {\bibfnamefont {B.~F.}\ \bibnamefont {Lasinski}}, \bibinfo {author}
  {\bibfnamefont {A.}~\bibnamefont {Lei}}, \bibinfo {author} {\bibfnamefont
  {A.~J.}\ \bibnamefont {MacKinnon}}, \bibinfo {author} {\bibfnamefont
  {P.}~\bibnamefont {Patel}}, \bibinfo {author} {\bibfnamefont
  {R.}~\bibnamefont {Stephens}}, \bibinfo {author} {\bibfnamefont
  {M.}~\bibnamefont {Tampo}}, \bibinfo {author} {\bibfnamefont {K.~A.}\
  \bibnamefont {Tanaka}}, \bibinfo {author} {\bibfnamefont {R.}~\bibnamefont
  {Town}}, \bibinfo {author} {\bibfnamefont {Y.}~\bibnamefont {Toyama}},
  \bibinfo {author} {\bibfnamefont {T.}~\bibnamefont {Tsutsumi}}, \bibinfo
  {author} {\bibfnamefont {S.~C.}\ \bibnamefont {Wilks}}, \bibinfo {author}
  {\bibfnamefont {T.}~\bibnamefont {Yabuuchi}}, \ and\ \bibinfo {author}
  {\bibfnamefont {J.}~\bibnamefont {Zheng}},\ }\href {\doibase
  10.1063/1.2774001} {\bibfield  {journal} {\bibinfo  {journal} {Phys.
  Plasmas}\ }\textbf {\bibinfo {volume} {14}},\ \bibinfo {pages} {092703}
  (\bibinfo {year} {2007})}\BibitemShut {NoStop}%
\bibitem [{\citenamefont {Borghesi}\ \emph {et~al.}(2004)\citenamefont
  {Borghesi}, \citenamefont {Mackinnon}, \citenamefont {Campbell},
  \citenamefont {Hicks}, \citenamefont {Kar}, \citenamefont {Patel},
  \citenamefont {Price}, \citenamefont {Romagnani}, \citenamefont {Schiavi},\
  and\ \citenamefont {Willi}}]{Borghesi2004}%
  \BibitemOpen
  \bibfield  {author} {\bibinfo {author} {\bibfnamefont {M.}~\bibnamefont
  {Borghesi}}, \bibinfo {author} {\bibfnamefont {A.}~\bibnamefont {Mackinnon}},
  \bibinfo {author} {\bibfnamefont {D.}~\bibnamefont {Campbell}}, \bibinfo
  {author} {\bibfnamefont {D.}~\bibnamefont {Hicks}}, \bibinfo {author}
  {\bibfnamefont {S.}~\bibnamefont {Kar}}, \bibinfo {author} {\bibfnamefont
  {P.}~\bibnamefont {Patel}}, \bibinfo {author} {\bibfnamefont
  {D.}~\bibnamefont {Price}}, \bibinfo {author} {\bibfnamefont
  {L.}~\bibnamefont {Romagnani}}, \bibinfo {author} {\bibfnamefont
  {A.}~\bibnamefont {Schiavi}}, \ and\ \bibinfo {author} {\bibfnamefont
  {O.}~\bibnamefont {Willi}},\ }\href {\doibase 10.1103/PhysRevLett.92.055003}
  {\bibfield  {journal} {\bibinfo  {journal} {Phys. Rev. Lett.}\ }\textbf
  {\bibinfo {volume} {92}},\ \bibinfo {pages} {055003} (\bibinfo {year}
  {2004})}\BibitemShut {NoStop}%
\bibitem [{\citenamefont {Roth}\ \emph {et~al.}(2001)\citenamefont {Roth},
  \citenamefont {Cowan}, \citenamefont {Key}, \citenamefont {Hatchett},
  \citenamefont {Brown}, \citenamefont {Fountain}, \citenamefont {Johnson},
  \citenamefont {Pennington}, \citenamefont {Snavely}, \citenamefont {Wilks},
  \citenamefont {Yasuike}, \citenamefont {Ruhl}, \citenamefont {Pegoraro},
  \citenamefont {Bulanov}, \citenamefont {Campbell}, \citenamefont {Perry},\
  and\ \citenamefont {Powell}}]{Roth2001}%
  \BibitemOpen
  \bibfield  {author} {\bibinfo {author} {\bibfnamefont {M.}~\bibnamefont
  {Roth}}, \bibinfo {author} {\bibfnamefont {T.}~\bibnamefont {Cowan}},
  \bibinfo {author} {\bibfnamefont {M.}~\bibnamefont {Key}}, \bibinfo {author}
  {\bibfnamefont {S.}~\bibnamefont {Hatchett}}, \bibinfo {author}
  {\bibfnamefont {C.}~\bibnamefont {Brown}}, \bibinfo {author} {\bibfnamefont
  {W.}~\bibnamefont {Fountain}}, \bibinfo {author} {\bibfnamefont
  {J.}~\bibnamefont {Johnson}}, \bibinfo {author} {\bibfnamefont
  {D.}~\bibnamefont {Pennington}}, \bibinfo {author} {\bibfnamefont
  {R.}~\bibnamefont {Snavely}}, \bibinfo {author} {\bibfnamefont
  {S.}~\bibnamefont {Wilks}}, \bibinfo {author} {\bibfnamefont
  {K.}~\bibnamefont {Yasuike}}, \bibinfo {author} {\bibfnamefont
  {H.}~\bibnamefont {Ruhl}}, \bibinfo {author} {\bibfnamefont {F.}~\bibnamefont
  {Pegoraro}}, \bibinfo {author} {\bibfnamefont {S.}~\bibnamefont {Bulanov}},
  \bibinfo {author} {\bibfnamefont {E.}~\bibnamefont {Campbell}}, \bibinfo
  {author} {\bibfnamefont {M.}~\bibnamefont {Perry}}, \ and\ \bibinfo {author}
  {\bibfnamefont {H.}~\bibnamefont {Powell}},\ }\href {\doibase
  10.1103/PhysRevLett.86.436} {\bibfield  {journal} {\bibinfo  {journal} {Phys.
  Rev. Lett.}\ }\textbf {\bibinfo {volume} {86}},\ \bibinfo {pages} {436}
  (\bibinfo {year} {2001})}\BibitemShut {NoStop}%
\bibitem [{\citenamefont {Hegelich}\ \emph
  {et~al.}(2013{\natexlab{a}})\citenamefont {Hegelich}, \citenamefont {Jung},
  \citenamefont {Albright}, \citenamefont {Cheung}, \citenamefont {Dromey},
  \citenamefont {Gautier}, \citenamefont {Hamilton}, \citenamefont {Letzring},
  \citenamefont {Munchhausen}, \citenamefont {Palaniyappan}, \citenamefont
  {Shah}, \citenamefont {Wu}, \citenamefont {Yin},\ and\ \citenamefont
  {Fernández}}]{Hegelich2013a}%
  \BibitemOpen
  \bibfield  {author} {\bibinfo {author} {\bibfnamefont {B.~M.}\ \bibnamefont
  {Hegelich}}, \bibinfo {author} {\bibfnamefont {D.}~\bibnamefont {Jung}},
  \bibinfo {author} {\bibfnamefont {B.~J.}\ \bibnamefont {Albright}}, \bibinfo
  {author} {\bibfnamefont {M.}~\bibnamefont {Cheung}}, \bibinfo {author}
  {\bibfnamefont {B.}~\bibnamefont {Dromey}}, \bibinfo {author} {\bibfnamefont
  {D.~C.}\ \bibnamefont {Gautier}}, \bibinfo {author} {\bibfnamefont
  {C.}~\bibnamefont {Hamilton}}, \bibinfo {author} {\bibfnamefont
  {S.}~\bibnamefont {Letzring}}, \bibinfo {author} {\bibfnamefont
  {R.}~\bibnamefont {Munchhausen}}, \bibinfo {author} {\bibfnamefont
  {S.}~\bibnamefont {Palaniyappan}}, \bibinfo {author} {\bibfnamefont
  {R.}~\bibnamefont {Shah}}, \bibinfo {author} {\bibfnamefont {H.-C.}\
  \bibnamefont {Wu}}, \bibinfo {author} {\bibfnamefont {L.}~\bibnamefont
  {Yin}}, \ and\ \bibinfo {author} {\bibfnamefont {J.~C.}\ \bibnamefont
  {Fernández}},\ }\href@noop {} {\enquote {\bibinfo {title} {160 {M}e{V}
  laser-accelerated protons from {CH}2 nanotargets for proton cancer
  therapy},}\ }\bibinfo {howpublished} {arXiv:1310.8650} (\bibinfo {year}
  {2013}{\natexlab{a}})\BibitemShut {NoStop}%
\bibitem [{\citenamefont {Jung}\ \emph
  {et~al.}(2013{\natexlab{a}})\citenamefont {Jung}, \citenamefont {Yin},
  \citenamefont {Gautier}, \citenamefont {Wu}, \citenamefont {Letzring},
  \citenamefont {Dromey}, \citenamefont {Shah}, \citenamefont {Palaniyappan},
  \citenamefont {Shimada}, \citenamefont {Johnson}, \citenamefont {Schreiber},
  \citenamefont {Habs}, \citenamefont {Fernandez}, \citenamefont {Hegelich},\
  and\ \citenamefont {Albright}}]{Jung2013a}%
  \BibitemOpen
  \bibfield  {author} {\bibinfo {author} {\bibfnamefont {D.}~\bibnamefont
  {Jung}}, \bibinfo {author} {\bibfnamefont {L.}~\bibnamefont {Yin}}, \bibinfo
  {author} {\bibfnamefont {D.~C.}\ \bibnamefont {Gautier}}, \bibinfo {author}
  {\bibfnamefont {H.~C.}\ \bibnamefont {Wu}}, \bibinfo {author} {\bibfnamefont
  {S.}~\bibnamefont {Letzring}}, \bibinfo {author} {\bibfnamefont
  {B.}~\bibnamefont {Dromey}}, \bibinfo {author} {\bibfnamefont
  {R.}~\bibnamefont {Shah}}, \bibinfo {author} {\bibfnamefont {S.}~\bibnamefont
  {Palaniyappan}}, \bibinfo {author} {\bibfnamefont {T.}~\bibnamefont
  {Shimada}}, \bibinfo {author} {\bibfnamefont {R.~P.}\ \bibnamefont
  {Johnson}}, \bibinfo {author} {\bibfnamefont {J.}~\bibnamefont {Schreiber}},
  \bibinfo {author} {\bibfnamefont {D.}~\bibnamefont {Habs}}, \bibinfo {author}
  {\bibfnamefont {J.~C.}\ \bibnamefont {Fernandez}}, \bibinfo {author}
  {\bibfnamefont {B.~M.}\ \bibnamefont {Hegelich}}, \ and\ \bibinfo {author}
  {\bibfnamefont {B.~J.}\ \bibnamefont {Albright}},\ }\href {\doibase
  10.1063/1.4817287} {\bibfield  {journal} {\bibinfo  {journal} {Phys.
  Plasmas}\ }\textbf {\bibinfo {volume} {20}},\ \bibinfo {pages} {083103}
  (\bibinfo {year} {2013}{\natexlab{a}})}\BibitemShut {NoStop}%
\bibitem [{\citenamefont {Yin}\ \emph {et~al.}(2006)\citenamefont {Yin},
  \citenamefont {Albright}, \citenamefont {Hegelich},\ and\ \citenamefont
  {Fernandez}}]{Yin2006}%
  \BibitemOpen
  \bibfield  {author} {\bibinfo {author} {\bibfnamefont {L.}~\bibnamefont
  {Yin}}, \bibinfo {author} {\bibfnamefont {B.~J.}\ \bibnamefont {Albright}},
  \bibinfo {author} {\bibfnamefont {B.~M.}\ \bibnamefont {Hegelich}}, \ and\
  \bibinfo {author} {\bibfnamefont {J.~C.}\ \bibnamefont {Fernandez}},\ }\href
  {\doibase 10.1017/S0263034606060459} {\bibfield  {journal} {\bibinfo
  {journal} {Laser Part. Beams}\ }\textbf {\bibinfo {volume} {24}},\ \bibinfo
  {pages} {291} (\bibinfo {year} {2006})}\BibitemShut {NoStop}%
\bibitem [{\citenamefont {Albright}\ \emph {et~al.}(2007)\citenamefont
  {Albright}, \citenamefont {Yin}, \citenamefont {Bowers}, \citenamefont
  {Hegelich}, \citenamefont {Flippo}, \citenamefont {Kwan},\ and\ \citenamefont
  {Fernandez}}]{Albright2007}%
  \BibitemOpen
  \bibfield  {author} {\bibinfo {author} {\bibfnamefont {B.~J.}\ \bibnamefont
  {Albright}}, \bibinfo {author} {\bibfnamefont {L.}~\bibnamefont {Yin}},
  \bibinfo {author} {\bibfnamefont {K.~J.}\ \bibnamefont {Bowers}}, \bibinfo
  {author} {\bibfnamefont {B.~M.}\ \bibnamefont {Hegelich}}, \bibinfo {author}
  {\bibfnamefont {K.~A.}\ \bibnamefont {Flippo}}, \bibinfo {author}
  {\bibfnamefont {T.~J.~T.}\ \bibnamefont {Kwan}}, \ and\ \bibinfo {author}
  {\bibfnamefont {J.~C.}\ \bibnamefont {Fernandez}},\ }\href {\doibase
  10.1063/1.2768933} {\bibfield  {journal} {\bibinfo  {journal} {Phys.
  Plasmas}\ }\textbf {\bibinfo {volume} {14}},\ \bibinfo {pages} {094502}
  (\bibinfo {year} {2007})}\BibitemShut {NoStop}%
\bibitem [{\citenamefont {Yin}\ \emph {et~al.}(2007)\citenamefont {Yin},
  \citenamefont {Albright}, \citenamefont {Hegelich}, \citenamefont {Bowers},
  \citenamefont {Flippo}, \citenamefont {Kwan},\ and\ \citenamefont
  {Fernandez}}]{Yin2007}%
  \BibitemOpen
  \bibfield  {author} {\bibinfo {author} {\bibfnamefont {L.}~\bibnamefont
  {Yin}}, \bibinfo {author} {\bibfnamefont {B.~J.}\ \bibnamefont {Albright}},
  \bibinfo {author} {\bibfnamefont {B.~M.}\ \bibnamefont {Hegelich}}, \bibinfo
  {author} {\bibfnamefont {K.~J.}\ \bibnamefont {Bowers}}, \bibinfo {author}
  {\bibfnamefont {K.~A.}\ \bibnamefont {Flippo}}, \bibinfo {author}
  {\bibfnamefont {T.~J.~T.}\ \bibnamefont {Kwan}}, \ and\ \bibinfo {author}
  {\bibfnamefont {J.~C.}\ \bibnamefont {Fernandez}},\ }\href {\doibase
  10.1063/1.2436857} {\bibfield  {journal} {\bibinfo  {journal} {Phys.
  Plasmas}\ }\textbf {\bibinfo {volume} {14}},\ \bibinfo {pages} {056706}
  (\bibinfo {year} {2007})}\BibitemShut {NoStop}%
\bibitem [{\citenamefont {Yin}\ \emph {et~al.}(2011)\citenamefont {Yin},
  \citenamefont {Albright}, \citenamefont {Bowers}, \citenamefont {Jung},
  \citenamefont {Fernandez},\ and\ \citenamefont {Hegelich}}]{Yin2011}%
  \BibitemOpen
  \bibfield  {author} {\bibinfo {author} {\bibfnamefont {L.}~\bibnamefont
  {Yin}}, \bibinfo {author} {\bibfnamefont {B.~J.}\ \bibnamefont {Albright}},
  \bibinfo {author} {\bibfnamefont {K.~J.}\ \bibnamefont {Bowers}}, \bibinfo
  {author} {\bibfnamefont {D.}~\bibnamefont {Jung}}, \bibinfo {author}
  {\bibfnamefont {J.~C.}\ \bibnamefont {Fernandez}}, \ and\ \bibinfo {author}
  {\bibfnamefont {B.~M.}\ \bibnamefont {Hegelich}},\ }\href {\doibase
  10.1103/PhysRevLett.107.045003} {\bibfield  {journal} {\bibinfo  {journal}
  {Phys. Rev. Lett.}\ }\textbf {\bibinfo {volume} {107}},\ \bibinfo {pages}
  {045003} (\bibinfo {year} {2011})}\BibitemShut {NoStop}%
\bibitem [{\citenamefont {Rus}\ \emph {et~al.}(2013)\citenamefont {Rus},
  \citenamefont {Bakule}, \citenamefont {Kramer}, \citenamefont {Korn},
  \citenamefont {Green}, \citenamefont {Novak}, \citenamefont {Fibrich},
  \citenamefont {Batysta}, \citenamefont {Thoma}, \citenamefont {Naylon},
  \citenamefont {Mazanec}, \citenamefont {Vitek}, \citenamefont {Barros},
  \citenamefont {Koutris}, \citenamefont {Hrebicek}, \citenamefont {Polan},
  \citenamefont {Base}, \citenamefont {Homer}, \citenamefont {Koselja},
  \citenamefont {Havlicek}, \citenamefont {Honsa}, \citenamefont {Novak},
  \citenamefont {Zervos}, \citenamefont {Korous}, \citenamefont {Laub},\ and\
  \citenamefont {Houzvicka}}]{Rus2013}%
  \BibitemOpen
  \bibfield  {author} {\bibinfo {author} {\bibfnamefont {B.}~\bibnamefont
  {Rus}}, \bibinfo {author} {\bibfnamefont {P.}~\bibnamefont {Bakule}},
  \bibinfo {author} {\bibfnamefont {D.}~\bibnamefont {Kramer}}, \bibinfo
  {author} {\bibfnamefont {G.}~\bibnamefont {Korn}}, \bibinfo {author}
  {\bibfnamefont {J.~T.}\ \bibnamefont {Green}}, \bibinfo {author}
  {\bibfnamefont {J.}~\bibnamefont {Novak}}, \bibinfo {author} {\bibfnamefont
  {M.}~\bibnamefont {Fibrich}}, \bibinfo {author} {\bibfnamefont
  {F.}~\bibnamefont {Batysta}}, \bibinfo {author} {\bibfnamefont
  {J.}~\bibnamefont {Thoma}}, \bibinfo {author} {\bibfnamefont
  {J.}~\bibnamefont {Naylon}}, \bibinfo {author} {\bibfnamefont
  {T.}~\bibnamefont {Mazanec}}, \bibinfo {author} {\bibfnamefont
  {M.}~\bibnamefont {Vitek}}, \bibinfo {author} {\bibfnamefont
  {R.}~\bibnamefont {Barros}}, \bibinfo {author} {\bibfnamefont
  {E.}~\bibnamefont {Koutris}}, \bibinfo {author} {\bibfnamefont
  {J.}~\bibnamefont {Hrebicek}}, \bibinfo {author} {\bibfnamefont
  {J.}~\bibnamefont {Polan}}, \bibinfo {author} {\bibfnamefont
  {R.}~\bibnamefont {Base}}, \bibinfo {author} {\bibfnamefont {P.}~\bibnamefont
  {Homer}}, \bibinfo {author} {\bibfnamefont {M.}~\bibnamefont {Koselja}},
  \bibinfo {author} {\bibfnamefont {T.}~\bibnamefont {Havlicek}}, \bibinfo
  {author} {\bibfnamefont {A.}~\bibnamefont {Honsa}}, \bibinfo {author}
  {\bibfnamefont {M.}~\bibnamefont {Novak}}, \bibinfo {author} {\bibfnamefont
  {C.}~\bibnamefont {Zervos}}, \bibinfo {author} {\bibfnamefont
  {P.}~\bibnamefont {Korous}}, \bibinfo {author} {\bibfnamefont
  {M.}~\bibnamefont {Laub}}, \ and\ \bibinfo {author} {\bibfnamefont
  {J.}~\bibnamefont {Houzvicka}},\ }in\ \href {\doibase 10.1117/12.2021264}
  {\emph {\bibinfo {booktitle} {HIGH-POWER, HIGH-ENERGY, AND HIGH-INTENSITY
  LASER TECHNOLOGY; AND RESEARCH USING EXTREME LIGHT: ENTERING NEW FRONTIERS
  WITH PETAWATT-CLASS LASERS}}},\ \bibinfo {series} {Proceedings of SPIE},
  Vol.\ \bibinfo {volume} {8780},\ \bibinfo {editor} {edited by\ \bibinfo
  {editor} {\bibfnamefont {J.}~\bibnamefont {Hein}}, \bibinfo {editor}
  {\bibfnamefont {G.}~\bibnamefont {Korn}}, \ and\ \bibinfo {editor}
  {\bibfnamefont {L.}~\bibnamefont {Silva}}}\ (\bibinfo {year} {2013})\ p.\
  \bibinfo {pages} {87801T}\BibitemShut {NoStop}%
\bibitem [{\citenamefont {Hegelich}\ \emph
  {et~al.}(2013{\natexlab{b}})\citenamefont {Hegelich}, \citenamefont
  {Pomerantz}, \citenamefont {Yin}, \citenamefont {Wu}, \citenamefont {Jung},
  \citenamefont {Albright}, \citenamefont {Gautier}, \citenamefont {Letzring},
  \citenamefont {Palaniyappan}, \citenamefont {Shah}, \citenamefont {Allinger},
  \citenamefont {Hoerlein}, \citenamefont {Schreiber}, \citenamefont {Habs},
  \citenamefont {Blakeney}, \citenamefont {Dyer}, \citenamefont {Fuller},
  \citenamefont {Gaul}, \citenamefont {Mccary}, \citenamefont {Meadows},
  \citenamefont {Wang}, \citenamefont {Ditmire},\ and\ \citenamefont
  {Fernandez}}]{Hegelich2013b}%
  \BibitemOpen
  \bibfield  {author} {\bibinfo {author} {\bibfnamefont {B.~M.}\ \bibnamefont
  {Hegelich}}, \bibinfo {author} {\bibfnamefont {I.}~\bibnamefont {Pomerantz}},
  \bibinfo {author} {\bibfnamefont {L.}~\bibnamefont {Yin}}, \bibinfo {author}
  {\bibfnamefont {H.~C.}\ \bibnamefont {Wu}}, \bibinfo {author} {\bibfnamefont
  {D.}~\bibnamefont {Jung}}, \bibinfo {author} {\bibfnamefont {B.~J.}\
  \bibnamefont {Albright}}, \bibinfo {author} {\bibfnamefont {D.~C.}\
  \bibnamefont {Gautier}}, \bibinfo {author} {\bibfnamefont {S.}~\bibnamefont
  {Letzring}}, \bibinfo {author} {\bibfnamefont {S.}~\bibnamefont
  {Palaniyappan}}, \bibinfo {author} {\bibfnamefont {R.}~\bibnamefont {Shah}},
  \bibinfo {author} {\bibfnamefont {K.}~\bibnamefont {Allinger}}, \bibinfo
  {author} {\bibfnamefont {R.}~\bibnamefont {Hoerlein}}, \bibinfo {author}
  {\bibfnamefont {J.}~\bibnamefont {Schreiber}}, \bibinfo {author}
  {\bibfnamefont {D.}~\bibnamefont {Habs}}, \bibinfo {author} {\bibfnamefont
  {J.}~\bibnamefont {Blakeney}}, \bibinfo {author} {\bibfnamefont
  {G.}~\bibnamefont {Dyer}}, \bibinfo {author} {\bibfnamefont {L.}~\bibnamefont
  {Fuller}}, \bibinfo {author} {\bibfnamefont {E.}~\bibnamefont {Gaul}},
  \bibinfo {author} {\bibfnamefont {E.}~\bibnamefont {Mccary}}, \bibinfo
  {author} {\bibfnamefont {A.~R.}\ \bibnamefont {Meadows}}, \bibinfo {author}
  {\bibfnamefont {C.}~\bibnamefont {Wang}}, \bibinfo {author} {\bibfnamefont
  {T.}~\bibnamefont {Ditmire}}, \ and\ \bibinfo {author} {\bibfnamefont
  {J.~C.}\ \bibnamefont {Fernandez}},\ }\href {\doibase
  10.1088/1367-2630/15/8/085015} {\bibfield  {journal} {\bibinfo  {journal}
  {New J. Phys.}\ }\textbf {\bibinfo {volume} {15}},\ \bibinfo {pages} {085015}
  (\bibinfo {year} {2013}{\natexlab{b}})}\BibitemShut {NoStop}%
\bibitem [{\citenamefont {Jung}\ \emph
  {et~al.}(2013{\natexlab{b}})\citenamefont {Jung}, \citenamefont {Yin},
  \citenamefont {Albright}, \citenamefont {Gautier}, \citenamefont {Letzring},
  \citenamefont {Dromey}, \citenamefont {Yeung}, \citenamefont {Hoerlein},
  \citenamefont {Shah}, \citenamefont {Palaniyappan}, \citenamefont {Allinger},
  \citenamefont {Schreiber}, \citenamefont {Bowers}, \citenamefont {Wu},
  \citenamefont {Fernandez}, \citenamefont {Habs},\ and\ \citenamefont
  {Hegelich}}]{Jung2013b}%
  \BibitemOpen
  \bibfield  {author} {\bibinfo {author} {\bibfnamefont {D.}~\bibnamefont
  {Jung}}, \bibinfo {author} {\bibfnamefont {L.}~\bibnamefont {Yin}}, \bibinfo
  {author} {\bibfnamefont {B.~J.}\ \bibnamefont {Albright}}, \bibinfo {author}
  {\bibfnamefont {D.~C.}\ \bibnamefont {Gautier}}, \bibinfo {author}
  {\bibfnamefont {S.}~\bibnamefont {Letzring}}, \bibinfo {author}
  {\bibfnamefont {B.}~\bibnamefont {Dromey}}, \bibinfo {author} {\bibfnamefont
  {M.}~\bibnamefont {Yeung}}, \bibinfo {author} {\bibfnamefont
  {R.}~\bibnamefont {Hoerlein}}, \bibinfo {author} {\bibfnamefont
  {R.}~\bibnamefont {Shah}}, \bibinfo {author} {\bibfnamefont {S.}~\bibnamefont
  {Palaniyappan}}, \bibinfo {author} {\bibfnamefont {K.}~\bibnamefont
  {Allinger}}, \bibinfo {author} {\bibfnamefont {J.}~\bibnamefont {Schreiber}},
  \bibinfo {author} {\bibfnamefont {K.~J.}\ \bibnamefont {Bowers}}, \bibinfo
  {author} {\bibfnamefont {H.-C.}\ \bibnamefont {Wu}}, \bibinfo {author}
  {\bibfnamefont {J.~C.}\ \bibnamefont {Fernandez}}, \bibinfo {author}
  {\bibfnamefont {D.}~\bibnamefont {Habs}}, \ and\ \bibinfo {author}
  {\bibfnamefont {B.~M.}\ \bibnamefont {Hegelich}},\ }\href {\doibase
  10.1088/1367-2630/15/2/023007} {\bibfield  {journal} {\bibinfo  {journal}
  {New J. Phys.}\ }\textbf {\bibinfo {volume} {15}},\ \bibinfo {pages} {023007}
  (\bibinfo {year} {2013}{\natexlab{b}})}\BibitemShut {NoStop}%
\bibitem [{\citenamefont {Yan}\ \emph {et~al.}(2010)\citenamefont {Yan},
  \citenamefont {Tajima}, \citenamefont {Hegelich}, \citenamefont {Yin},\ and\
  \citenamefont {Habs}}]{Yan2010}%
  \BibitemOpen
  \bibfield  {author} {\bibinfo {author} {\bibfnamefont {X.~Q.}\ \bibnamefont
  {Yan}}, \bibinfo {author} {\bibfnamefont {T.}~\bibnamefont {Tajima}},
  \bibinfo {author} {\bibfnamefont {M.}~\bibnamefont {Hegelich}}, \bibinfo
  {author} {\bibfnamefont {L.}~\bibnamefont {Yin}}, \ and\ \bibinfo {author}
  {\bibfnamefont {D.}~\bibnamefont {Habs}},\ }\href {\doibase
  10.1007/s00340-009-3707-5} {\bibfield  {journal} {\bibinfo  {journal} {Appl.
  Phys. B}\ }\textbf {\bibinfo {volume} {{98}}},\ \bibinfo {pages} {711}
  (\bibinfo {year} {2010})}\BibitemShut {NoStop}%
\bibitem [{\citenamefont {Sentoku}\ \emph {et~al.}(2003)\citenamefont
  {Sentoku}, \citenamefont {Cowan}, \citenamefont {Kemp},\ and\ \citenamefont
  {Ruhl}}]{Sentoku2003}%
  \BibitemOpen
  \bibfield  {author} {\bibinfo {author} {\bibfnamefont {Y.}~\bibnamefont
  {Sentoku}}, \bibinfo {author} {\bibfnamefont {T.}~\bibnamefont {Cowan}},
  \bibinfo {author} {\bibfnamefont {A.}~\bibnamefont {Kemp}}, \ and\ \bibinfo
  {author} {\bibfnamefont {H.}~\bibnamefont {Ruhl}},\ }\href {\doibase
  10.1063/1.1556298} {\bibfield  {journal} {\bibinfo  {journal} {Phys.
  Plasmas}\ }\textbf {\bibinfo {volume} {10}},\ \bibinfo {pages} {2009}
  (\bibinfo {year} {2003})}\BibitemShut {NoStop}%
\bibitem [{\citenamefont {Qiao}\ \emph {et~al.}(2012)\citenamefont {Qiao},
  \citenamefont {Kar}, \citenamefont {Geissler}, \citenamefont {Gibbon},
  \citenamefont {Zepf},\ and\ \citenamefont {Borghesi}}]{Qiao2012}%
  \BibitemOpen
  \bibfield  {author} {\bibinfo {author} {\bibfnamefont {B.}~\bibnamefont
  {Qiao}}, \bibinfo {author} {\bibfnamefont {S.}~\bibnamefont {Kar}}, \bibinfo
  {author} {\bibfnamefont {M.}~\bibnamefont {Geissler}}, \bibinfo {author}
  {\bibfnamefont {P.}~\bibnamefont {Gibbon}}, \bibinfo {author} {\bibfnamefont
  {M.}~\bibnamefont {Zepf}}, \ and\ \bibinfo {author} {\bibfnamefont
  {M.}~\bibnamefont {Borghesi}},\ }\href {\doibase
  10.1103/PhysRevLett.108.115002} {\bibfield  {journal} {\bibinfo  {journal}
  {Phys. Rev. Lett.}\ }\textbf {\bibinfo {volume} {108}},\ \bibinfo {pages}
  {115002} (\bibinfo {year} {2012})}\BibitemShut {NoStop}%
\bibitem [{\citenamefont {Gibbon}(2005)}]{Gibbon2005}%
  \BibitemOpen
  \bibfield  {author} {\bibinfo {author} {\bibfnamefont {P.}~\bibnamefont
  {Gibbon}},\ }\href@noop {} {\emph {\bibinfo {title} {Short pulse laser
  interactions with matter: An introduction}}}\ (\bibinfo  {publisher}
  {Imperial College Press},\ \bibinfo {year} {2005})\BibitemShut {NoStop}%
\bibitem [{\citenamefont {Lefebvre}\ and\ \citenamefont
  {Bonnaud}(1995)}]{Lefebvre1995}%
  \BibitemOpen
  \bibfield  {author} {\bibinfo {author} {\bibfnamefont {E.}~\bibnamefont
  {Lefebvre}}\ and\ \bibinfo {author} {\bibfnamefont {G.}~\bibnamefont
  {Bonnaud}},\ }\href {\doibase 10.1103/PhysRevLett.74.2002} {\bibfield
  {journal} {\bibinfo  {journal} {Phys. Rev. Lett.}\ }\textbf {\bibinfo
  {volume} {74}},\ \bibinfo {pages} {2002} (\bibinfo {year}
  {1995})}\BibitemShut {NoStop}%
\bibitem [{\citenamefont {Sakagami}\ and\ \citenamefont
  {Mima}(1996)}]{Sakagami1996}%
  \BibitemOpen
  \bibfield  {author} {\bibinfo {author} {\bibfnamefont {H.}~\bibnamefont
  {Sakagami}}\ and\ \bibinfo {author} {\bibfnamefont {K.}~\bibnamefont
  {Mima}},\ }\href {\doibase 10.1103/PhysRevE.54.1870} {\bibfield  {journal}
  {\bibinfo  {journal} {Phys. Rev. E}\ }\textbf {\bibinfo {volume} {54}},\
  \bibinfo {pages} {1870} (\bibinfo {year} {1996})}\BibitemShut {NoStop}%
\bibitem [{\citenamefont {Gordienko}\ and\ \citenamefont
  {Pukhov}(2005)}]{Gordienko2005}%
  \BibitemOpen
  \bibfield  {author} {\bibinfo {author} {\bibfnamefont {S.}~\bibnamefont
  {Gordienko}}\ and\ \bibinfo {author} {\bibfnamefont {A.}~\bibnamefont
  {Pukhov}},\ }\href {\doibase 10.1063/1.1884126} {\bibfield  {journal}
  {\bibinfo  {journal} {Phys. Plasmas}\ }\textbf {\bibinfo {volume} {12}},\
  \bibinfo {pages} {043109} (\bibinfo {year} {2005})}\BibitemShut {NoStop}%
\bibitem [{\citenamefont {Hillier}\ \emph {et~al.}(2013)\citenamefont
  {Hillier}, \citenamefont {Danson}, \citenamefont {Duffield}, \citenamefont
  {Egan}, \citenamefont {Elsmere}, \citenamefont {Girling}, \citenamefont
  {Harvey}, \citenamefont {Hopps}, \citenamefont {Norman}, \citenamefont
  {Parker}, \citenamefont {Treadwell}, \citenamefont {Winter},\ and\
  \citenamefont {Bett}}]{Hillier2013}%
  \BibitemOpen
  \bibfield  {author} {\bibinfo {author} {\bibfnamefont {D.}~\bibnamefont
  {Hillier}}, \bibinfo {author} {\bibfnamefont {C.}~\bibnamefont {Danson}},
  \bibinfo {author} {\bibfnamefont {S.}~\bibnamefont {Duffield}}, \bibinfo
  {author} {\bibfnamefont {D.}~\bibnamefont {Egan}}, \bibinfo {author}
  {\bibfnamefont {S.}~\bibnamefont {Elsmere}}, \bibinfo {author} {\bibfnamefont
  {M.}~\bibnamefont {Girling}}, \bibinfo {author} {\bibfnamefont
  {E.}~\bibnamefont {Harvey}}, \bibinfo {author} {\bibfnamefont
  {N.}~\bibnamefont {Hopps}}, \bibinfo {author} {\bibfnamefont
  {M.}~\bibnamefont {Norman}}, \bibinfo {author} {\bibfnamefont
  {S.}~\bibnamefont {Parker}}, \bibinfo {author} {\bibfnamefont
  {P.}~\bibnamefont {Treadwell}}, \bibinfo {author} {\bibfnamefont
  {D.}~\bibnamefont {Winter}}, \ and\ \bibinfo {author} {\bibfnamefont
  {T.}~\bibnamefont {Bett}},\ }\href {\doibase 10.1364/AO.52.004258} {\bibfield
   {journal} {\bibinfo  {journal} {Appl. Optics}\ }\textbf {\bibinfo {volume}
  {52}},\ \bibinfo {pages} {4258} (\bibinfo {year} {2013})}\BibitemShut
  {NoStop}%
\bibitem [{\citenamefont {Thaury}\ \emph {et~al.}(2007)\citenamefont {Thaury},
  \citenamefont {Quere}, \citenamefont {Geindre}, \citenamefont {Levy},
  \citenamefont {Ceccotti}, \citenamefont {Monot}, \citenamefont {Bougeard},
  \citenamefont {Reau}, \citenamefont {D'Oliveira}, \citenamefont {Audebert},
  \citenamefont {Marjoribanks},\ and\ \citenamefont {Martin}}]{Thaury2007}%
  \BibitemOpen
  \bibfield  {author} {\bibinfo {author} {\bibfnamefont {C.}~\bibnamefont
  {Thaury}}, \bibinfo {author} {\bibfnamefont {F.}~\bibnamefont {Quere}},
  \bibinfo {author} {\bibfnamefont {J.-P.}\ \bibnamefont {Geindre}}, \bibinfo
  {author} {\bibfnamefont {A.}~\bibnamefont {Levy}}, \bibinfo {author}
  {\bibfnamefont {T.}~\bibnamefont {Ceccotti}}, \bibinfo {author}
  {\bibfnamefont {P.}~\bibnamefont {Monot}}, \bibinfo {author} {\bibfnamefont
  {M.}~\bibnamefont {Bougeard}}, \bibinfo {author} {\bibfnamefont
  {F.}~\bibnamefont {Reau}}, \bibinfo {author} {\bibfnamefont {P.}~\bibnamefont
  {D'Oliveira}}, \bibinfo {author} {\bibfnamefont {P.}~\bibnamefont
  {Audebert}}, \bibinfo {author} {\bibfnamefont {R.}~\bibnamefont
  {Marjoribanks}}, \ and\ \bibinfo {author} {\bibfnamefont {P.~H.}\
  \bibnamefont {Martin}},\ }\href {\doibase 10.1038/nphys595} {\bibfield
  {journal} {\bibinfo  {journal} {Nature Phys.}\ }\textbf {\bibinfo {volume}
  {3}},\ \bibinfo {pages} {424} (\bibinfo {year} {2007})}\BibitemShut {NoStop}%
\bibitem [{\citenamefont {Psikal}\ \emph {et~al.}(2006)\citenamefont {Psikal},
  \citenamefont {Limpouch}, \citenamefont {Kawata},\ and\ \citenamefont
  {Andreev}}]{Psikal2006}%
  \BibitemOpen
  \bibfield  {author} {\bibinfo {author} {\bibfnamefont {J.}~\bibnamefont
  {Psikal}}, \bibinfo {author} {\bibfnamefont {J.}~\bibnamefont {Limpouch}},
  \bibinfo {author} {\bibfnamefont {S.}~\bibnamefont {Kawata}}, \ and\ \bibinfo
  {author} {\bibfnamefont {A.~A.}\ \bibnamefont {Andreev}},\ }\href {\doibase
  10.1007/s10582-006-0246-8} {\bibfield  {journal} {\bibinfo  {journal} {Czech.
  J. Phys.}\ }\textbf {\bibinfo {volume} {56}},\ \bibinfo {pages} {B515}
  (\bibinfo {year} {2006})}\BibitemShut {NoStop}%
\bibitem [{\citenamefont {Ueda}\ \emph {et~al.}(1994)\citenamefont {Ueda},
  \citenamefont {Omura}, \citenamefont {Matsumoto},\ and\ \citenamefont
  {Okuzawa}}]{Ueda1994}%
  \BibitemOpen
  \bibfield  {author} {\bibinfo {author} {\bibfnamefont {H.}~\bibnamefont
  {Ueda}}, \bibinfo {author} {\bibfnamefont {Y.}~\bibnamefont {Omura}},
  \bibinfo {author} {\bibfnamefont {H.}~\bibnamefont {Matsumoto}}, \ and\
  \bibinfo {author} {\bibfnamefont {T.}~\bibnamefont {Okuzawa}},\ }\href
  {\doibase 10.1016/0010-4655(94)90071-X} {\bibfield  {journal} {\bibinfo
  {journal} {Comp. Phys. Comm.}\ }\textbf {\bibinfo {volume} {79}},\ \bibinfo
  {pages} {249} (\bibinfo {year} {1994})}\BibitemShut {NoStop}%
\bibitem [{\citenamefont {Limpouch}\ \emph {et~al.}(2013)\citenamefont
  {Limpouch}, \citenamefont {Psikal}, \citenamefont {Klimo}, \citenamefont
  {Vyskocil}, \citenamefont {Proska}, \citenamefont {Novotny}, \citenamefont
  {Stolcova},\ and\ \citenamefont {Kveton}}]{Limpouch2013}%
  \BibitemOpen
  \bibfield  {author} {\bibinfo {author} {\bibfnamefont {J.}~\bibnamefont
  {Limpouch}}, \bibinfo {author} {\bibfnamefont {J.}~\bibnamefont {Psikal}},
  \bibinfo {author} {\bibfnamefont {O.}~\bibnamefont {Klimo}}, \bibinfo
  {author} {\bibfnamefont {J.}~\bibnamefont {Vyskocil}}, \bibinfo {author}
  {\bibfnamefont {J.}~\bibnamefont {Proska}}, \bibinfo {author} {\bibfnamefont
  {F.}~\bibnamefont {Novotny}}, \bibinfo {author} {\bibfnamefont
  {L.}~\bibnamefont {Stolcova}}, \ and\ \bibinfo {author} {\bibfnamefont
  {M.}~\bibnamefont {Kveton}},\ }in\ \href {\doibase 10.1117/12.2021848} {\emph
  {\bibinfo {booktitle} {HIGH-POWER, HIGH-ENERGY, AND HIGH-INTENSITY LASER
  TECHNOLOGY; AND RESEARCH USING EXTREME LIGHT: ENTERING NEW FRONTIERS WITH
  PETAWATT-CLASS LASERS}}},\ \bibinfo {series} {Proceedings of SPIE}, Vol.\
  \bibinfo {volume} {8780},\ \bibinfo {editor} {edited by\ \bibinfo {editor}
  {\bibfnamefont {J.}~\bibnamefont {Hein}}, \bibinfo {editor} {\bibfnamefont
  {G.}~\bibnamefont {Korn}}, \ and\ \bibinfo {editor} {\bibfnamefont
  {L.}~\bibnamefont {Silva}}}\ (\bibinfo {year} {2013})\ p.\ \bibinfo {pages}
  {878027}\BibitemShut {NoStop}%
\bibitem [{\citenamefont {Snavely}\ \emph {et~al.}(2000)\citenamefont
  {Snavely}, \citenamefont {Key}, \citenamefont {Hatchett}, \citenamefont
  {Cowan}, \citenamefont {Roth}, \citenamefont {Phillips}, \citenamefont
  {Stoyer}, \citenamefont {Henry}, \citenamefont {Sangster}, \citenamefont
  {Singh}, \citenamefont {Wilks}, \citenamefont {MacKinnon}, \citenamefont
  {Offenberger}, \citenamefont {Pennington}, \citenamefont {Yasuike},
  \citenamefont {Langdon}, \citenamefont {Lasinski}, \citenamefont {Johnson},
  \citenamefont {Perry},\ and\ \citenamefont {Campbell}}]{Snavely2000}%
  \BibitemOpen
  \bibfield  {author} {\bibinfo {author} {\bibfnamefont {R.}~\bibnamefont
  {Snavely}}, \bibinfo {author} {\bibfnamefont {M.}~\bibnamefont {Key}},
  \bibinfo {author} {\bibfnamefont {S.}~\bibnamefont {Hatchett}}, \bibinfo
  {author} {\bibfnamefont {T.}~\bibnamefont {Cowan}}, \bibinfo {author}
  {\bibfnamefont {M.}~\bibnamefont {Roth}}, \bibinfo {author} {\bibfnamefont
  {T.}~\bibnamefont {Phillips}}, \bibinfo {author} {\bibfnamefont
  {M.}~\bibnamefont {Stoyer}}, \bibinfo {author} {\bibfnamefont
  {E.}~\bibnamefont {Henry}}, \bibinfo {author} {\bibfnamefont
  {T.}~\bibnamefont {Sangster}}, \bibinfo {author} {\bibfnamefont
  {M.}~\bibnamefont {Singh}}, \bibinfo {author} {\bibfnamefont
  {S.}~\bibnamefont {Wilks}}, \bibinfo {author} {\bibfnamefont
  {A.}~\bibnamefont {MacKinnon}}, \bibinfo {author} {\bibfnamefont
  {A.}~\bibnamefont {Offenberger}}, \bibinfo {author} {\bibfnamefont
  {D.}~\bibnamefont {Pennington}}, \bibinfo {author} {\bibfnamefont
  {K.}~\bibnamefont {Yasuike}}, \bibinfo {author} {\bibfnamefont
  {A.}~\bibnamefont {Langdon}}, \bibinfo {author} {\bibfnamefont
  {B.}~\bibnamefont {Lasinski}}, \bibinfo {author} {\bibfnamefont
  {J.}~\bibnamefont {Johnson}}, \bibinfo {author} {\bibfnamefont
  {M.}~\bibnamefont {Perry}}, \ and\ \bibinfo {author} {\bibfnamefont
  {E.}~\bibnamefont {Campbell}},\ }\href {\doibase 10.1103/PhysRevLett.85.2945}
  {\bibfield  {journal} {\bibinfo  {journal} {Phys. Rev. Lett.}\ }\textbf
  {\bibinfo {volume} {85}},\ \bibinfo {pages} {2945} (\bibinfo {year}
  {2000})}\BibitemShut {NoStop}%
\bibitem [{\citenamefont {Wilks}\ \emph {et~al.}(2001)\citenamefont {Wilks},
  \citenamefont {Langdon}, \citenamefont {Cowan}, \citenamefont {Roth},
  \citenamefont {Singh}, \citenamefont {Hatchett}, \citenamefont {Key},
  \citenamefont {Pennington}, \citenamefont {MacKinnon},\ and\ \citenamefont
  {Snavely}}]{Wilks2001}%
  \BibitemOpen
  \bibfield  {author} {\bibinfo {author} {\bibfnamefont {S.}~\bibnamefont
  {Wilks}}, \bibinfo {author} {\bibfnamefont {A.}~\bibnamefont {Langdon}},
  \bibinfo {author} {\bibfnamefont {T.}~\bibnamefont {Cowan}}, \bibinfo
  {author} {\bibfnamefont {M.}~\bibnamefont {Roth}}, \bibinfo {author}
  {\bibfnamefont {M.}~\bibnamefont {Singh}}, \bibinfo {author} {\bibfnamefont
  {S.}~\bibnamefont {Hatchett}}, \bibinfo {author} {\bibfnamefont
  {M.}~\bibnamefont {Key}}, \bibinfo {author} {\bibfnamefont {D.}~\bibnamefont
  {Pennington}}, \bibinfo {author} {\bibfnamefont {A.}~\bibnamefont
  {MacKinnon}}, \ and\ \bibinfo {author} {\bibfnamefont {R.}~\bibnamefont
  {Snavely}},\ }\href {\doibase 10.1063/1.1333697} {\bibfield  {journal}
  {\bibinfo  {journal} {Phys. Plasmas}\ }\textbf {\bibinfo {volume} {8}},\
  \bibinfo {pages} {542} (\bibinfo {year} {2001})}\BibitemShut {NoStop}%
\bibitem [{\citenamefont {Pukhov}\ \emph {et~al.}(1999)\citenamefont {Pukhov},
  \citenamefont {Sheng},\ and\ \citenamefont {Meyer-ter Vehn}}]{Pukhov1999}%
  \BibitemOpen
  \bibfield  {author} {\bibinfo {author} {\bibfnamefont {A.}~\bibnamefont
  {Pukhov}}, \bibinfo {author} {\bibfnamefont {Z.}~\bibnamefont {Sheng}}, \
  and\ \bibinfo {author} {\bibfnamefont {J.}~\bibnamefont {Meyer-ter Vehn}},\
  }\href {\doibase 10.1063/1.873242} {\bibfield  {journal} {\bibinfo  {journal}
  {Phys. Plasmas}\ }\textbf {\bibinfo {volume} {6}},\ \bibinfo {pages} {2847}
  (\bibinfo {year} {1999})}\BibitemShut {NoStop}%
\bibitem [{\citenamefont {Neely}\ \emph {et~al.}(2006)\citenamefont {Neely},
  \citenamefont {Foster}, \citenamefont {Robinson}, \citenamefont {Lindau},
  \citenamefont {Lundh}, \citenamefont {Persson}, \citenamefont {Wahlstrom},\
  and\ \citenamefont {McKenna}}]{Neely2006}%
  \BibitemOpen
  \bibfield  {author} {\bibinfo {author} {\bibfnamefont {D.}~\bibnamefont
  {Neely}}, \bibinfo {author} {\bibfnamefont {P.}~\bibnamefont {Foster}},
  \bibinfo {author} {\bibfnamefont {A.}~\bibnamefont {Robinson}}, \bibinfo
  {author} {\bibfnamefont {F.}~\bibnamefont {Lindau}}, \bibinfo {author}
  {\bibfnamefont {O.}~\bibnamefont {Lundh}}, \bibinfo {author} {\bibfnamefont
  {A.}~\bibnamefont {Persson}}, \bibinfo {author} {\bibfnamefont {C.~G.}\
  \bibnamefont {Wahlstrom}}, \ and\ \bibinfo {author} {\bibfnamefont
  {P.}~\bibnamefont {McKenna}},\ }\href {\doibase 10.1063/1.2220011} {\bibfield
   {journal} {\bibinfo  {journal} {Appl. Phys. Lett.}\ }\textbf {\bibinfo
  {volume} {89}},\ \bibinfo {pages} {021502} (\bibinfo {year}
  {2006})}\BibitemShut {NoStop}%
\bibitem [{\citenamefont {Ceccotti}\ \emph {et~al.}(2007)\citenamefont
  {Ceccotti}, \citenamefont {Levy}, \citenamefont {Popescu}, \citenamefont
  {Reau}, \citenamefont {D'Oliveira}, \citenamefont {Monot}, \citenamefont
  {Geindre}, \citenamefont {Lefebvre},\ and\ \citenamefont
  {Martin}}]{Ceccotti2007}%
  \BibitemOpen
  \bibfield  {author} {\bibinfo {author} {\bibfnamefont {T.}~\bibnamefont
  {Ceccotti}}, \bibinfo {author} {\bibfnamefont {A.}~\bibnamefont {Levy}},
  \bibinfo {author} {\bibfnamefont {H.}~\bibnamefont {Popescu}}, \bibinfo
  {author} {\bibfnamefont {F.}~\bibnamefont {Reau}}, \bibinfo {author}
  {\bibfnamefont {P.}~\bibnamefont {D'Oliveira}}, \bibinfo {author}
  {\bibfnamefont {P.}~\bibnamefont {Monot}}, \bibinfo {author} {\bibfnamefont
  {J.~P.}\ \bibnamefont {Geindre}}, \bibinfo {author} {\bibfnamefont
  {E.}~\bibnamefont {Lefebvre}}, \ and\ \bibinfo {author} {\bibfnamefont
  {P.}~\bibnamefont {Martin}},\ }\href {\doibase 10.1103/PhysRevLett.99.185002}
  {\bibfield  {journal} {\bibinfo  {journal} {Phys. Rev. Lett.}\ }\textbf
  {\bibinfo {volume} {99}},\ \bibinfo {pages} {185002} (\bibinfo {year}
  {2007})}\BibitemShut {NoStop}%
\bibitem [{\citenamefont {Klimo}\ \emph {et~al.}(2011)\citenamefont {Klimo},
  \citenamefont {Psikal}, \citenamefont {Limpouch}, \citenamefont {Proska},
  \citenamefont {Novotny}, \citenamefont {Ceccotti}, \citenamefont {Floquet},\
  and\ \citenamefont {Kawata}}]{Klimo2011}%
  \BibitemOpen
  \bibfield  {author} {\bibinfo {author} {\bibfnamefont {O.}~\bibnamefont
  {Klimo}}, \bibinfo {author} {\bibfnamefont {J.}~\bibnamefont {Psikal}},
  \bibinfo {author} {\bibfnamefont {J.}~\bibnamefont {Limpouch}}, \bibinfo
  {author} {\bibfnamefont {J.}~\bibnamefont {Proska}}, \bibinfo {author}
  {\bibfnamefont {F.}~\bibnamefont {Novotny}}, \bibinfo {author} {\bibfnamefont
  {T.}~\bibnamefont {Ceccotti}}, \bibinfo {author} {\bibfnamefont
  {V.}~\bibnamefont {Floquet}}, \ and\ \bibinfo {author} {\bibfnamefont
  {S.}~\bibnamefont {Kawata}},\ }\href {\doibase 10.1088/1367-2630/13/5/053028}
  {\bibfield  {journal} {\bibinfo  {journal} {New J. Phys.}\ }\textbf {\bibinfo
  {volume} {13}},\ \bibinfo {pages} {053028} (\bibinfo {year}
  {2011})}\BibitemShut {NoStop}%
\bibitem [{\citenamefont {Margarone}\ \emph {et~al.}(2012)\citenamefont
  {Margarone}, \citenamefont {Klimo}, \citenamefont {Kim}, \citenamefont
  {Prokupek}, \citenamefont {Limpouch}, \citenamefont {Jeong}, \citenamefont
  {Mocek}, \citenamefont {Psikal}, \citenamefont {Kim}, \citenamefont {Proska},
  \citenamefont {Nam}, \citenamefont {Stolcova}, \citenamefont {Choi},
  \citenamefont {Lee}, \citenamefont {Sung}, \citenamefont {Yu},\ and\
  \citenamefont {Korn}}]{Margarone2012}%
  \BibitemOpen
  \bibfield  {author} {\bibinfo {author} {\bibfnamefont {D.}~\bibnamefont
  {Margarone}}, \bibinfo {author} {\bibfnamefont {O.}~\bibnamefont {Klimo}},
  \bibinfo {author} {\bibfnamefont {I.~J.}\ \bibnamefont {Kim}}, \bibinfo
  {author} {\bibfnamefont {J.}~\bibnamefont {Prokupek}}, \bibinfo {author}
  {\bibfnamefont {J.}~\bibnamefont {Limpouch}}, \bibinfo {author}
  {\bibfnamefont {T.~M.}\ \bibnamefont {Jeong}}, \bibinfo {author}
  {\bibfnamefont {T.}~\bibnamefont {Mocek}}, \bibinfo {author} {\bibfnamefont
  {J.}~\bibnamefont {Psikal}}, \bibinfo {author} {\bibfnamefont {H.~T.}\
  \bibnamefont {Kim}}, \bibinfo {author} {\bibfnamefont {J.}~\bibnamefont
  {Proska}}, \bibinfo {author} {\bibfnamefont {K.~H.}\ \bibnamefont {Nam}},
  \bibinfo {author} {\bibfnamefont {L.}~\bibnamefont {Stolcova}}, \bibinfo
  {author} {\bibfnamefont {I.~W.}\ \bibnamefont {Choi}}, \bibinfo {author}
  {\bibfnamefont {S.~K.}\ \bibnamefont {Lee}}, \bibinfo {author} {\bibfnamefont
  {J.~H.}\ \bibnamefont {Sung}}, \bibinfo {author} {\bibfnamefont {T.~J.}\
  \bibnamefont {Yu}}, \ and\ \bibinfo {author} {\bibfnamefont {G.}~\bibnamefont
  {Korn}},\ }\href {\doibase 10.1103/PhysRevLett.109.234801} {\bibfield
  {journal} {\bibinfo  {journal} {Phys. Rev. Lett.}\ }\textbf {\bibinfo
  {volume} {109}},\ \bibinfo {pages} {234801} (\bibinfo {year}
  {2012})}\BibitemShut {NoStop}%
\bibitem [{\citenamefont {Floquet}\ \emph {et~al.}(2013)\citenamefont
  {Floquet}, \citenamefont {Klimo}, \citenamefont {Psikal}, \citenamefont
  {Velyhan}, \citenamefont {Limpouch}, \citenamefont {Proska}, \citenamefont
  {Novotny}, \citenamefont {Stolcova}, \citenamefont {Macchi}, \citenamefont
  {Sgattoni}, \citenamefont {Vassura}, \citenamefont {Labate}, \citenamefont
  {Baffigi}, \citenamefont {Gizzi}, \citenamefont {Martin},\ and\ \citenamefont
  {Ceccotti}}]{Floquet2013}%
  \BibitemOpen
  \bibfield  {author} {\bibinfo {author} {\bibfnamefont {V.}~\bibnamefont
  {Floquet}}, \bibinfo {author} {\bibfnamefont {O.}~\bibnamefont {Klimo}},
  \bibinfo {author} {\bibfnamefont {J.}~\bibnamefont {Psikal}}, \bibinfo
  {author} {\bibfnamefont {A.}~\bibnamefont {Velyhan}}, \bibinfo {author}
  {\bibfnamefont {J.}~\bibnamefont {Limpouch}}, \bibinfo {author}
  {\bibfnamefont {J.}~\bibnamefont {Proska}}, \bibinfo {author} {\bibfnamefont
  {F.}~\bibnamefont {Novotny}}, \bibinfo {author} {\bibfnamefont
  {L.}~\bibnamefont {Stolcova}}, \bibinfo {author} {\bibfnamefont
  {A.}~\bibnamefont {Macchi}}, \bibinfo {author} {\bibfnamefont
  {A.}~\bibnamefont {Sgattoni}}, \bibinfo {author} {\bibfnamefont
  {L.}~\bibnamefont {Vassura}}, \bibinfo {author} {\bibfnamefont
  {L.}~\bibnamefont {Labate}}, \bibinfo {author} {\bibfnamefont
  {F.}~\bibnamefont {Baffigi}}, \bibinfo {author} {\bibfnamefont {L.~A.}\
  \bibnamefont {Gizzi}}, \bibinfo {author} {\bibfnamefont {P.}~\bibnamefont
  {Martin}}, \ and\ \bibinfo {author} {\bibfnamefont {T.}~\bibnamefont
  {Ceccotti}},\ }\href {\doibase 10.1063/1.4819239} {\bibfield  {journal}
  {\bibinfo  {journal} {J. Appl. Phys.}\ }\textbf {\bibinfo {volume} {114}},\
  \bibinfo {pages} {083305} (\bibinfo {year} {2013})}\BibitemShut {NoStop}%
\bibitem [{\citenamefont {Sgattoni}\ \emph {et~al.}(2012)\citenamefont
  {Sgattoni}, \citenamefont {Londrillo}, \citenamefont {Macchi},\ and\
  \citenamefont {Passoni}}]{Sgattoni2012}%
  \BibitemOpen
  \bibfield  {author} {\bibinfo {author} {\bibfnamefont {A.}~\bibnamefont
  {Sgattoni}}, \bibinfo {author} {\bibfnamefont {P.}~\bibnamefont {Londrillo}},
  \bibinfo {author} {\bibfnamefont {A.}~\bibnamefont {Macchi}}, \ and\ \bibinfo
  {author} {\bibfnamefont {M.}~\bibnamefont {Passoni}},\ }\href {\doibase
  10.1103/PhysRevE.85.036405} {\bibfield  {journal} {\bibinfo  {journal} {Phys.
  Rev. E}\ }\textbf {\bibinfo {volume} {85}},\ \bibinfo {pages} {036405}
  (\bibinfo {year} {2012})}\BibitemShut {NoStop}%
\bibitem [{\citenamefont {Ceccotti}\ \emph {et~al.}(2013)\citenamefont
  {Ceccotti}, \citenamefont {Floquet}, \citenamefont {Sgattoni}, \citenamefont
  {Bigongiari}, \citenamefont {Klimo}, \citenamefont {Raynaud}, \citenamefont
  {Riconda}, \citenamefont {Heron}, \citenamefont {Baffigi}, \citenamefont
  {Labate}, \citenamefont {Gizzi}, \citenamefont {Vassura}, \citenamefont
  {Fuchs}, \citenamefont {Passoni}, \citenamefont {Kveton}, \citenamefont
  {Novotny}, \citenamefont {Possolt}, \citenamefont {Prokupek}, \citenamefont
  {Proska}, \citenamefont {Psikal}, \citenamefont {Stolcova}, \citenamefont
  {Velyhan}, \citenamefont {Bougeard}, \citenamefont {D'Oliveira},
  \citenamefont {Tcherbakoff}, \citenamefont {Reau}, \citenamefont {Martin},\
  and\ \citenamefont {Macchi}}]{Ceccotti2013}%
  \BibitemOpen
  \bibfield  {author} {\bibinfo {author} {\bibfnamefont {T.}~\bibnamefont
  {Ceccotti}}, \bibinfo {author} {\bibfnamefont {V.}~\bibnamefont {Floquet}},
  \bibinfo {author} {\bibfnamefont {A.}~\bibnamefont {Sgattoni}}, \bibinfo
  {author} {\bibfnamefont {A.}~\bibnamefont {Bigongiari}}, \bibinfo {author}
  {\bibfnamefont {O.}~\bibnamefont {Klimo}}, \bibinfo {author} {\bibfnamefont
  {M.}~\bibnamefont {Raynaud}}, \bibinfo {author} {\bibfnamefont
  {C.}~\bibnamefont {Riconda}}, \bibinfo {author} {\bibfnamefont
  {A.}~\bibnamefont {Heron}}, \bibinfo {author} {\bibfnamefont
  {F.}~\bibnamefont {Baffigi}}, \bibinfo {author} {\bibfnamefont
  {L.}~\bibnamefont {Labate}}, \bibinfo {author} {\bibfnamefont {L.~A.}\
  \bibnamefont {Gizzi}}, \bibinfo {author} {\bibfnamefont {L.}~\bibnamefont
  {Vassura}}, \bibinfo {author} {\bibfnamefont {J.}~\bibnamefont {Fuchs}},
  \bibinfo {author} {\bibfnamefont {M.}~\bibnamefont {Passoni}}, \bibinfo
  {author} {\bibfnamefont {M.}~\bibnamefont {Kveton}}, \bibinfo {author}
  {\bibfnamefont {F.}~\bibnamefont {Novotny}}, \bibinfo {author} {\bibfnamefont
  {M.}~\bibnamefont {Possolt}}, \bibinfo {author} {\bibfnamefont
  {J.}~\bibnamefont {Prokupek}}, \bibinfo {author} {\bibfnamefont
  {J.}~\bibnamefont {Proska}}, \bibinfo {author} {\bibfnamefont
  {J.}~\bibnamefont {Psikal}}, \bibinfo {author} {\bibfnamefont
  {L.}~\bibnamefont {Stolcova}}, \bibinfo {author} {\bibfnamefont
  {A.}~\bibnamefont {Velyhan}}, \bibinfo {author} {\bibfnamefont
  {M.}~\bibnamefont {Bougeard}}, \bibinfo {author} {\bibfnamefont
  {P.}~\bibnamefont {D'Oliveira}}, \bibinfo {author} {\bibfnamefont
  {O.}~\bibnamefont {Tcherbakoff}}, \bibinfo {author} {\bibfnamefont
  {F.}~\bibnamefont {Reau}}, \bibinfo {author} {\bibfnamefont {P.}~\bibnamefont
  {Martin}}, \ and\ \bibinfo {author} {\bibfnamefont {A.}~\bibnamefont
  {Macchi}},\ }\href {\doibase 10.1103/PhysRevLett.111.185001} {\bibfield
  {journal} {\bibinfo  {journal} {Phys. Rev. Lett.}\ }\textbf {\bibinfo
  {volume} {111}},\ \bibinfo {pages} {185001} (\bibinfo {year}
  {2013})}\BibitemShut {NoStop}%
\end{thebibliography}

\providecommand{\noopsort}[1]{}\providecommand{\singleletter}[1]{#1}%

\end{document}